\documentclass[amsmath,twocolumn,amssymb,floatfix,showpacs,
superscriptaddress,nofootinbib,longbibliography]{revtex4-1}

\usepackage{graphicx}
\usepackage{bm}
\usepackage{braket}
\usepackage[dvipsnames]{xcolor}
\usepackage{subfigure}
\usepackage{tikz}
\usepackage{pgffor}
\usepackage[colorlinks=true,linktoc=page,
linkcolor=BrickRed,citecolor=magenta,urlcolor=purple]{hyperref}
\usepackage{enumitem}
\usepackage[normalem]{ulem}  

\mathchardef\mhyphen="2D

\newcommand{\etal}{{\it et al.}}

\newcommand\bea{\begin{eqnarray}}
\newcommand\eea{\end{eqnarray}}
\newcommand\beq{\begin{equation}}
\newcommand\eeq{\end{equation}}

\definecolor{lime}{HTML}{A6CE39}
\DeclareRobustCommand{\orcidicon}{\hspace{-1.0mm}
\begin{tikzpicture}
 		\draw[lime, fill=lime] (0.0,0.0) circle [radius=0.15]
 		node[white] {{\fontfamily{qag}\selectfont \tiny \,ID}};
		\draw[white, fill=white] (-0.0525,0.095) circle [radius=0.007];
 	\end{tikzpicture}
 	\hspace{-3.0mm}
 }
 \foreach \x in {A, ..., Z}{\expandafter\xdef\csname orcid\x\endcsname{%
 	\noexpand\href{https://orcid.org/\csname orcidauthor\x\endcsname}{\noexpand\orcidicon}}}

\AtBeginDocument{%
    \newwrite\bibnotes
    \def\bibnotesext{Notes.bib}
    \immediate\openout\bibnotes=\jobname\bibnotesext
    \immediate\write\bibnotes{@CONTROL{REVTEX41Control}}
    \immediate\write\bibnotes{@CONTROL{%
    apsrev41Control,author="08",editor="1",pages="1",title="0",year="1"}}
     \if@filesw
     \immediate\write\@auxout{\string\citation{apsrev41Control}}%
    \fi
}%
\begin{document}

\title{Floquet-engineered diode performance in a Majorana-quantum dot Josephson junction}

\author{Koustav Roy}
\email{koustav.roy@iitg.ac.in}
\affiliation{Department of Physics, Indian Institute of Technology Guwahati-Guwahati, 781039 Assam, India}

\author{Gourab Paul\orcidB{}}
\email{p.gourab@iitg.ac.in}
\affiliation{Department of Physics, Indian Institute of Technology Guwahati-Guwahati, 781039 Assam, India}

\author{Debika Debnath\orcidC{}}
\email{debika@prl.res.in (corresponding author)}
\affiliation{Theoretical Physics Division, Physical Research Laboratory, Navrangpura, Ahmedabad-380009, India}

\author{Kuntal Bhattacharyya\orcidD{}}
\email{kuntalphy@iitg.ac.in}
\affiliation{Department of Physics, Indian Institute of Technology Guwahati-Guwahati, 781039 Assam, India}

\author{Saurabh Basu}
\email{saurabh@iitg.ac.in}
\affiliation{Department of Physics, Indian Institute of Technology Guwahati-Guwahati, 781039 Assam, India}

\begin{abstract}
We study nonreciprocal signatures of Josephson current in a quantum dot (QD)-based Josephson junction that comprises of two periodically driven Kitaev chains (KCs) coupled with an intervening QD. The simultaneous breaking of the inversion symmetry ($\mathcal{IS}$) and the time-reversal symmetry ($\mathcal{TRS}$), indispensable for the Josephson diode effect (JDE), is achieved solely via the two Floquet drives that differ by a finite phase, which eventually results in a nonreciprocal current and hence yields a finite JDE. It may be noted that the Floquet Majorana modes generated at both the far ends of the KCs (away from the QD) and adjacent to the QD junctions mediate the Josephson current owing to a finite superconducting (SC) phase difference in the two KCs. We calculate the time-averaged Josephson current and inspect the tunability of the current-phase relation (CPR) to ascertain the diode characteristics. The asymmetric Floquet drive also manifests an anomalous Josephson current signature in our KC-QD-KC Josephson junction. Furthermore, additional control over the QD energy level can be achieved via an external gate voltage that renders flexibility for the diode to act as an SC switching device. Tuning different system parameters, such as the chemical potential of the KCs, Floquet frequency, the relative phase mismatch of the drives, and the gate voltage, our model shows a maximum rectification to be around $70\%$. Summarizing, our study provides an alternative scenario, replacing the traditional usage of an external magnetic field and spin-orbit coupling effects in a Josephson diode via asymmetrically driven Kitaev leads that entail Majorana-mediated transport.
\end{abstract}

\maketitle
\section{Introduction}\label{intro}
Unidirectional current flow through semiconductor-based 
p-n junctions~\cite{Braun1875, Shockley1950} had shaped the quantum technologies and device fabrication for decades, until there was a pronounced shift towards the advent of SC heterojunctions manifesting the direction-dependent dissipationless current~\cite{Ando2020}. Moreover, the phase difference-induced supercurrent through the SC junctions nowadays is at the forefront of modern condensed matter research due to its feasible nonreciprocal signature which leads to the superconducting diode effect~\cite{Strambini2022,Narita2022,Daido2022,Picoli2023, Hu2023,Zinkl2022,Banerjee2024,Yerin2024,Meyer2024, Legg2022}. 
In 2018, Tokura and Nagaosa~\cite{Nagaosa2018} conjectured the possibility of achieving nonreciprocal supercurrent in a noncentrosymmetric crystal through an external magnetic field or magneteochiral anisotropy. The first experimental realization of the superconducting diode effect by Ando \etal~\cite{Ando2020} in an artificial Rashba superconductor has opened up a compelling avenue for understanding its underlying mechanism based on the principle of simultaneous breaking of the intrinsic symmetries, such as the $\mathcal{IS}$ and the $\mathcal{TRS}$.
Building on their experiment, several theoretical propositions have been put forward in the literature that suggest symmetry breaking through magnetochiral anisotropy~\cite{Baumgartner2022, Bauriedl2022, Legg2022} or intrinsic magnetic moment~\cite{Debnath2024b} that breaks the $\mathcal{TRS}$. Otherwise, an external magnetic field~\cite{Ando2020, Sun2023, Debnath2024a, Soori2025, Meyer2024} fulfills the requirement.

Experimental realization of the superconducting diode effect through an intermediate weak link has shown the possibility of controlling the direction of SC current in the Josephson junction~\cite{Josephson1962} which results in an externally controllable JDE~\cite{Misaki2021, Zhang2022, Souto2022, Wei2022, Davydova2022, Fominov2022, Lu2023, Huang2024, Chatterjee2024, Cheng2023, Steiner2023, Debnath2024a, Fracassi2024, Debnath2024b, Zalom2024} with different critical currents in the forward and reverse bias conditions, satisfying $I_{c}(\phi)\neq -I_{c}(-\phi)$.
Corresponding to a forward bias, the maximum critical current appears as $I_{c}^{+}$ for the SC phase difference $0 \le \phi \le \pi$. For the reverse direction, where $\pi \le \phi \le 2\pi$ (or $-\pi \le \phi \le 0$), the maximum critical current is $I_{c}^{-}$. The sign change in $I_{c}^{+}$ and $I_{c}^{-}$ occurs due to an asymmetric free energy ($E(\phi) \ne E(-\phi)$) which is only possible to achieve via simultaneous breaking of $\mathcal{IS}$ and $\mathcal{TRS}$. Hence, the finite difference between these two critical Josephson currents $\Delta I_{c}=I_{c}^{+}-|I_{c}^{-}|$ generates nonreciprocity, which eventually determines the efficiency of the Josephson junction, quantified in the literature through the rectification factor (RF),  $\mathcal{R}$~\cite{Davydova2022, Liu2024, Cayao2024, Debnath2024b} as
\begin{equation}
    \mathcal{R} = \left( \frac{\Delta I_{c}}{I_c^+ + |I_c^-|} \right) \times 100 \%.
    \label{diode_RF}
\end{equation}

Since the first evidence~\cite{Ando2020} of SC diode effect, numerous studies have proposed the JDE across a wide range of systems which includes Rashba superconductors~\cite{Bauriedl2022, Ando2020}, van der Waals heterostructures~\cite{Wu2022}, topological insulators~\cite{Legg2022}, Dirac semimetals~\cite{Chen2023, Yu2024}, altermagnets~\cite{Banerjee2024, Chakraborty2024} etc. The JDE has also been observed in systems such as a single magnetic atom~\cite{Sun2023, Trahms2023}, carbon nanotube~\cite{Nagaosa2023}, InSb nanoflag~\cite{Turini2022}, normal metals~\cite{Liu2024a}, and band-asymmetric metals~\cite{Soori2023}. Additionally, topological superconductors~\cite{Cayao2024, Liu2024b} have promised potential applicability as a JDE, highlighting its versatility spanning over different classes of materials.

In particular, recent reports on the JDE have garnered the role of a QD~\cite{Alivisatos1996,Folsch2014,Zwerver2022,Burkard2023} that exhibits a single quantized energy level, and can act as a quantum point contact as the simplest prototype of a weak link for a Josephson junction setup~\cite{ Cheng2023, Sun2023, Taberner2023, Trahms2023, Debnath2024a, Debnath2024b}. In addition to its fundamental interest, the connection of an external gate voltage to the QD energy level effectively tunes the barrier potential created across the QD junction~\cite{Alivisatos1996,Zwerver2022,Burkard2023,Mayer2020, Miller2003}, which manifests a control over the nonreciprocity of Josephson current~\cite{Gupta2023,Yan2025, Mayer2020}. Gupta \etal~\cite{Gupta2023} have studied such effects in a three-terminal Josephson device using an external electrostatic gating. Moreover, in an InAs nanosheet-based Josephson junction, suppression of the gate voltage dominates the current nonreciprocity~\cite{Yan2025, Mayer2020}.  Along the same lines, theoretical approaches also predict the control over the unidirectional Josephson current upon tuning the QD energy across a Josephson junction~\cite{Cheng2023, Debnath2024a, Debnath2024b}, making it a noteworthy switching device. The anatomy of these properties of the QD has built the QD-based diode as an efficient tool for studying the quantum devices starting from quantum spin-qubits~\cite{Loss1998, Fominov2022, Cuozzo2024, Greco2024, Yu2024}, transmon circuit~\cite{Bargerbos2022}, quantum interferometers~\cite{Souto2022, Ciaccia2023}, etc. 

Of late, studies have advocated JDE in a driven Josephson junction~\cite{Taberner2023, Soori2023b,Scheer2025}. Tarberner \etal~\cite{Taberner2023} have shown the appearance of an anomalous Josephson current in a driven double QD-Josephson junction leading to the CPR: $I (\phi=0) \neq 0$, which is also explained as a $\phi_0$-Josephson junction~\cite{Mayer2020, Davydova2022, Martin2009, Yokoyama2014, Assouline2019} in the literature. The appearance of this finite Josephson current in the absence of any external SC phase bias causes an extra phase shift in the Josephson CPR as $ I(\phi_0)=0$ (i.e., the Josephson current will become zero at $\phi_0 \ne 0)$ which has been a consequence of the spin-orbit coupling and a magnetic field in a Josephson junction~\cite{Buzdin2008, Krive2005, Reynoso2008, Martin2009, Yokoyama2014}. The findings of anomalous Josephson current have set a new horizon to facilitate nonreciprocity in the context of Floquet-driven systems~\cite{Taberner2023, Soori2023b}. In Ref.~\cite{Taberner2023}, the Floquet drive is well-studied for the driven double QD Josephson junction, albeit without the exploration of the Majorana quasiparticles. Whereas, the other study~\cite{Soori2023b} shows the effect of two driven KCs coupled with each other without any intermediate weak link, leaving large possibilities to explore in detail.

The appearance of the Majorana end modes, namely the Majorana zero-modes (MZMs) and the (Floquet) Majorana $\pi$-modes (MPMs)~\cite{Kundu2013,Peng2021,Liu2019,Wang2024,Kumari2024} has validated the possibility of distinct topological quantum phases with different types of Majoranas~\cite{Alicea2012,Beenakker2013,Leijnse2012} and exploiting their characteristics in topological Josephson junction~\cite{Kundu2013,Cayao2017,Baldo2023,Liu2018,Liu2024b,Cayao2024,Steiner2023,Gao2024,Taberner2023,Peng2021,Liu2019,Wang2024,Kumari2024,Escribano2025}. Influenced by Kitaev’s seminal work on one-dimensional ($1$D) model of a spinless $p$-wave superconductor~\cite{Kitaev2001} describing two spatially separated Majorana modes, the driven KC~\cite{Tong2013, Thakurathi2014, Wang2019, Soori2023b, Roy2024, Roy2024_rashba, Lutchyn2010, Kumari2024,Wu2023} has unfolded significant attributes. Despite the exotic nature of $p$-wave superconductors, observable signatures, such as zero-bias conductance peaks \cite{Law2009,Das2012,Churchill2013,Finck2013,Liu2012,Deng2012} lead to unique nontrivial transport phenomena.
Recently, within a similar genre, Kumari~\etal~\cite{Kumari2024} have examined the behaviour of the Josephson current due to the unpaired Floquet-Majorana bound states in a Josephson junction using the Floquet-Keldysh sum rule for the driven superconductors and concluded with the robust predictions for a driven topological system. 
All these studies on topological Josephson junction substantially enable the detection and control of the Majorana-bound states with great precision~\cite{Nayak2008,DasSarma2005}. Although Majorana-based JDE~\cite{Liu2024b,Cayao2024,Steiner2023,Gao2024,Mondal2025} have been reported earlier, the diode signatures in those are induced by a magnetic field and spin-orbit coupling. Thus, the evidence of such a diode simply with a Floquet drive is scarce~\cite{Cayssol2013, Runder2013, Leon2013, Runder2020}, which particularly motivates our study.

The rationale behind conceiving the present investigation is multifold. Firstly, the role of the Floquet drive has not been established fully as an alternate tool to stimulate the JDE, except for a few proposals~\cite{Taberner2023,Soori2023b}. Secondly, though QD-based weak links between multiple KCs have been investigated in topological transport~\cite{Li2014,Taberner2023,Liu2018,Medina-Cuy2023}, achieving the JDE has mainly focused on driving the QDs~\cite{Taberner2023}, instead of driving the KCs by operating them out of equilibrium, which could reveal novel phases (which are unattainable in static configurations). Furthermore, such an approach considering a driven QD lacks a clear physical realization of Majoranas, as the driving term effectively plays the role of a spin-orbit coupling in the QD rather than introducing new topological features. Thus, a driven QD does not alter the fundamental topological characteristics of the KC, thereby failing to generate the MPMs, which are essential signatures of Floquet-induced topology \cite{Cayssol2013, Runder2013, Leon2013, Runder2020}. 
\begin{figure*}
\begin{center}
\includegraphics[width=0.6\linewidth]{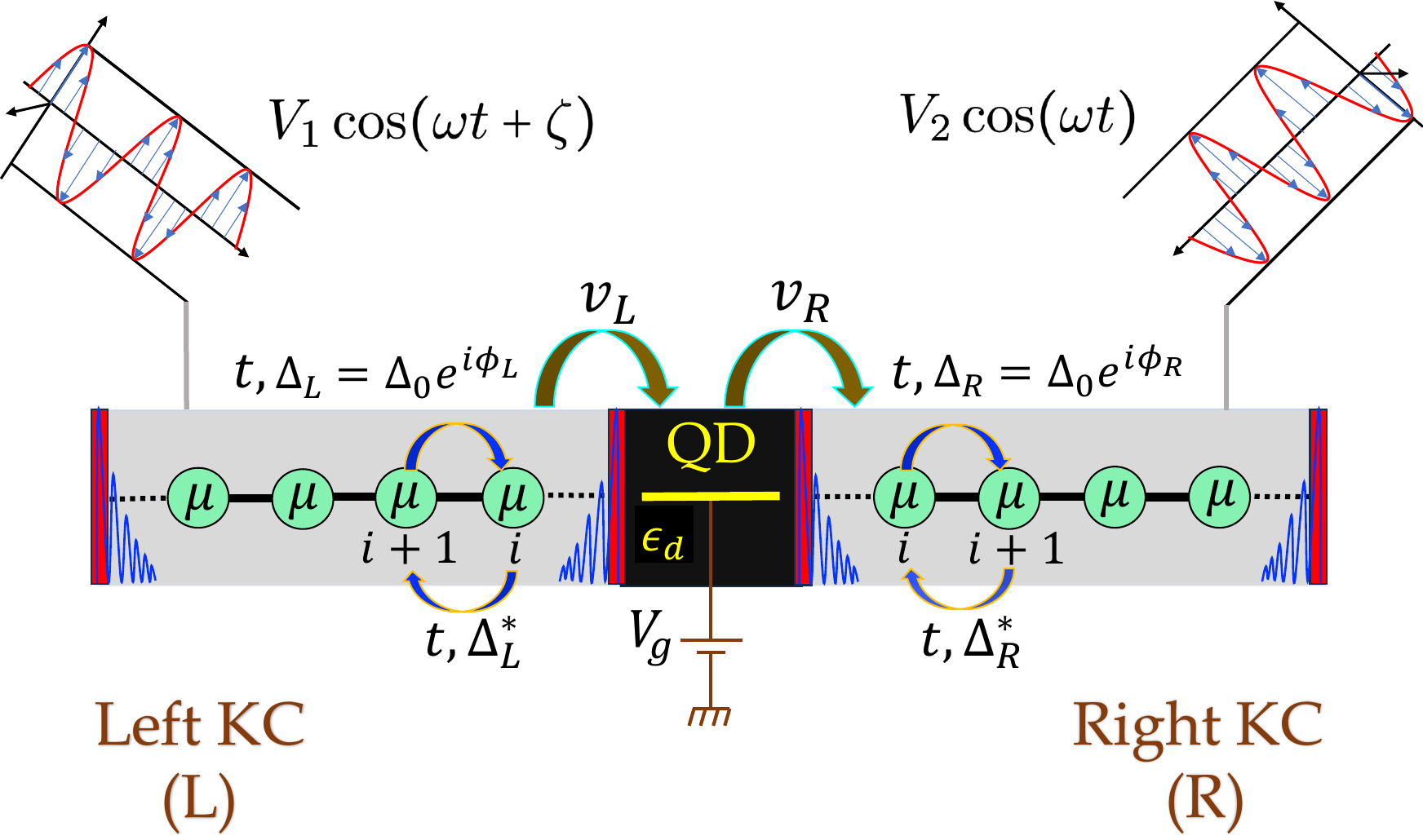}
\caption{A schematic illustration of a Josephson junction composed of two $p$-wave spinless KCs with the chemical potential, $\mu$, nearest-neighbour hopping strength, $t$, and the SC pairing potentials, \(\Delta_L\) and \(\Delta_R\) respectively for the left (L) and right (R) KCs connected via a QD-based weak link (highlighted in black) is presented. The SC phase difference, \(\phi_L - \phi_R = \phi\), is maintained across the junction. The tunnelling amplitudes are denoted as \(v_L\) and \(v_R\), corresponding to the L-QD and QD-R segments, respectively. The energy level of the QD, $\epsilon_d$ is tuned via an external gate voltage $V_g$. The Floquet version of this model is realized under the application of two periodic drives applied to the L and R KCs with a Floquet frequency, $\omega$, accompanied by a finite phase difference, $\zeta$. The Majorana modes are localized at the left and right edges of the KCs and also at the interfaces of the QD, which are denoted as red bars with their confined probability densities marked in blue lines.}  
\label{fig:model}
\end{center}

\end{figure*}
To the best of our knowledge, the hitherto proposals of the JDE (both Majorana and QD-based) are achieved either in the presence of both an external magnetic field and Rashba spin-orbit interaction (for breaking the $\mathcal{TRS}$ and $\mathcal{IS}$ simultaneously), or with driven QDs. However, the search for a field-free JDE still continues. In contrast, driven KCs demonstrate not only the enhancement of their inherent topological properties but also introduce \textit{artificial} spin-orbit coupling and Zeeman terms (playing the role of the external agents), which could result in richer nontrivial transport phenomena. Therefore, applying a Floquet drive to the KCs (keeping the QD as an undriven weak link) may be the simplest approach to achieve a topologically enriched JDE.

The scarcity of such an example with driven KCs intrigues the following questions:
\vspace{-2mm}
\begin{enumerate}[label=(\roman*)]
\setlength{\itemsep}{0pt}
\setlength{\parskip}{0pt}
\item Can a QD placed in between two driven KCs act as a \textit{magnetic and spin-orbit} field-free Josephson diode?
\item Does the application of a Floquet drive to the SC leads made of two $p$-wave KCs suffice to break $\mathcal{TRS}$ and $\mathcal{IS}$ for achieving a finite rectification of the diode?
\item How indispensable is the tuning of the RF upon varying the Floquet drive parameters and energy level of the QD?
\end{enumerate}

With this motivation, we consider two periodically driven KCs coupled with an intermediate weak link of a QD (schematically represented in Fig.~\ref{fig:model}), where the description of this generic model is discussed in the caption of the figure. Furthermore, inducing an asymmetric phase factor between the two Floquet drives associated with the two KCs may break $\mathcal{TRS}$ and $\mathcal{IS}$, which in turn results in a finite nonreciprocity and rectification fulfilling the appropriate requirement of the diode effect. 

We organize the remainder of the paper as follows. In Sec.~\ref{static}, we present the static version of our model in which we introduce the static Hamiltonian in Sec.~\ref{static H model} and briefly discuss the essential physics of the MZMs and subsequently define the static Josephson current in Sec.~\ref{static current}. Sec.~\ref{driven} is devoted to the Floquet scenario hosting additional MPMs, which establishes the requirement of an asymmetric drive in breaking $\mathcal{TRS}$ and $\mathcal{IS}$ and generating a nonreciprocal current. Thereafter, we elaborately discuss the numerical results for the driven case, which includes the behaviour of the driven Josephson current, its CPR characteristics in Sec.~\ref{driven JC}, and the effects of the drive parameters on controlling the nonreciprocity of the driven Josephson current in Sec.~\ref{driven JC2}, and examine the diode's efficiency in terms of the rectification factor in Sec.~\ref{RF} both quantitatively and qualitatively. Sec.~\ref{gate tunability} explains the gate tunability of the diode effect. Finally, we summarize and conclude our findings in Sec.~\ref{summary}.


\section{Static KC-QD-KC Josephson junction hosting MZMs}\label{static}
In our model, we consider a semiconductor QD as the barrier (weak link), which is sandwiched between two finite spinless $1$D $p$-wave KCs~\cite{Kitaev2001,Lutchyn2010, Alicea2010, Das2012, Lang2012,Oreg2010} (acting as the SC leads), labeled as L (R) for the left (right) lead, forming a prototype for the Josephson junction (as shown in Fig.\,\ref{fig:model}). The static version of our model should be visualized without the drives that are depicted via the cosine modulations in Fig.\,\ref{fig:model}. Experimental fabrication of the QD~\cite{Alivisatos1996,Folsch2014,Zwerver2022,Burkard2023,Gupta2023} can be achieved by considering a single atom or single molecular structure~\cite{Folsch2014} or by stacking different semiconducting materials of uneven thicknesses between a source and a drain, producing a large potential barrier in a quantum well-like heterostructure~\cite{Alivisatos1996,Zwerver2022,Burkard2023,Gupta2023}. In general, a QD can possess multiple quantized levels in which a given number of electrons can be accommodated because of this strong quantum confinement effect. However, in our study, we assume that the confinement effect is strong enough to cause sufficiently large level spacing. Hence, the QD can be viewed as a single spinful orbital, which is sufficiently relevant for the current to flow across the Josephson junction. Furthermore, as said earlier, one may note that the discrete energy level of the QD is tunable by an external gate voltage $V_g$ that may influence the transport through the QD. Therefore, the effect of $V_g$ should be considered while deciphering the role of the QD in such a Josephson junction setup. 
\subsection{Static Hamiltonian}\label{static H model}
Before delving into the scenario of a periodically driven Josephson junction, we introduce the static (non-driven) Hamiltonian for our system that mimics a normal Josephson junction and can be formulated as
\begin{eqnarray}
\mathcal{H}_{\text{stat}} &=& \mathcal{H}_{\text{L}}+\mathcal{H}_{\text{R}}+\mathcal{H}_{\text{QD}}+\mathcal{H}_{\text{T}},\label{Eq2}
\end{eqnarray}
where the components of $\mathcal{H}_{\text{stat}}$ are expressed as
\begin{eqnarray}
\mathcal{H}_{\text{L(R)}} &=& -\mu \sum_{j=1}^{N_s} c_{L(R),j}^{\dagger} c_{L(R),j} \nonumber\\ 
&& -t \sum_{j=1}^{N_s-1}\left[c_{L(R),j}^{\dagger}c_{L(R),j+1}+\text{h.c.}\right]\nonumber\\
&& +\sum_{j=1}^{N_s-1} \left[\Delta_{L(R)} c_{L(R),j+1}^{\dagger} c_{L(R),j}^{\dagger} + \text{h.c.}\right],
\label{Ham:L,R}\\
\mathcal{H}_{\text{QD}}&=&\sum_{\sigma} (\epsilon_{d} - eV_{g} ) d^{\dagger}_{\sigma}d_{\sigma},
\label{Ham:QD}\\
\mathcal{H}_{\text{T}} &=& v_L c_{L,1}^{\dagger} d_{\sigma} + v_R  c_{R,1}^{\dagger} d_{\sigma} + \text{h.c.},
\label{Ham:tunnel}
\end{eqnarray}
which respectively represent the Hamiltonian for the left (right) KC (Eq.~\eqref{Ham:L,R}), the central QD (Eq.~\eqref{Ham:QD}), and the tunnelling between KCs and the QD (Eq.~\eqref{Ham:tunnel}). In Eq.~\eqref{Ham:L,R}, the first term of $\mathcal{H}_{\text{L(R)}}$ denotes the onsite energy for the left (right) KC with $\mu$ being the chemical potential, and $c_{L(R),j}^{\dagger}$($c_{L(R),j}$) being the creation (annihilation) operator of a spinless electron for the $j$-th site of each $p$-wave KC, while the second term signifies the nearest-neighbour (NN) hopping between two adjacent sites of the KC with a coupling strength $t$. $N_s$ represents the total number of sites for each KC. The SC pairing potential $\Delta_{L(R)}$ of the left (right) KC is designated by the third term of Eq.~\eqref{Ham:L,R} which reads as $\Delta_{L(R)} = \Delta_0 e^{i\phi_{L(R)}}$, where $\Delta_0$ and $\phi_{L(R)}$  denote the SC order parameter and the corresponding phase pertaining to the left (right) KCs, respectively. In Eq.~\eqref{Ham:QD}, $\epsilon_d$ stands for a single quantized energy level of the QD, which can be tuned externally by a gate voltage $V_g$~\cite{Mayer2020, Miller2003,Alivisatos1996,Zwerver2022,Burkard2023,Gupta2023}. $d^{\dagger}_{\sigma}$($d_{\sigma}$) represents the creation (annihilation) operator for an electron present in the QD corresponding to the spin state, $\sigma$ ($\uparrow$ or $\downarrow$). It can be seen from Fig.~\,\ref{fig:model} that the QD is coupled to the first ($j=1$) site of each KC on either side, and hence the tunneling amplitudes of the left KC-to-QD and QD-to-right KC channels are designated by $v_L$ and $v_R$, respectively in Eq.~\eqref{Ham:tunnel}.
%
\subsection{MZMs and static Josephson current}\label{static current}
At this point, it is crucial to understand the requisites of the $p$-wave KC and inspect the fate of the MZMs in such a QD-based heterojunction. We briefly present the numerical results of our static Josephson junction model in Appendix~\ref{appendixA}. In both the $\mathcal{TRS}$ and particle-hole preserved classes, for a finite $\Delta_0$, the Majorana operators $(c_{L(R),j}^{\dagger}+c_{L(R),j})$ and $i(c_{L(R),j}^{\dagger}-c_{L(R),j})$ synthesize robust MZMs at the edges under an open boundary condition provided the chemical potential ($\mu$) of the KC satisfies $-2t<\mu<2t$. 
In our study, the MZMs should appear across the junctions of the QD which can be analyzed by studying the behaviour of the energy spectrum and the static Josephson current flowing through such a KC-QD-KC type Josephson junction. 

At first, we employ a canonical transformation, $e^{\mathcal{S}}$ with a unitary operator $U$ to $\mathcal{H}_{\text{stat}}$, aiming to decouple the SC gap $\Delta_0$ from the SC phase $\phi_{L(R)}$. Such a recipe simplifies the Hamiltonian and makes it more tractable to diagnose the system's behaviour in the presence of the SC phase. The generator $\mathcal{S}$ that facilitates this transformation is given by~\cite{QSun2000a}
\begin{equation}
\mathcal{S} = \sum_{j=1}^{N_s}\left[\frac{i \phi_L}{2} c_{L,j}^{\dagger} c_{L,j} + \frac{i \phi_R}{2}  c_{R,j}^{\dagger} c_{R,j}\right].
\label{generator}
\end{equation}
The transformed Hamiltonian, namely $\tilde{\mathcal{H}}_{\text{stat}}=e^{\mathcal{S}} \mathcal{H}_{\text{stat}} e^{-{\mathcal{S}}}$ reads as follows:

\begin{eqnarray}
\tilde{\mathcal{H}}_{\text{stat}} &=& -\mu \sum_{\substack{\alpha\in L,R\\j=1}}^{N_s} c_{\alpha,j}^{\dagger} c_{\alpha,j} + \mathcal{H}_{\text{QD}}\nonumber\\
&&+\sum_{\substack{\alpha\in L,R\\j=1}}^{N_s-1} \left[-tc_{\alpha,j}^{\dagger}c_{\alpha,j+1}+\Delta_0 c_{\alpha,j+1}^{\dagger} c_{\alpha,j}^{\dagger} + H.c.\right]\nonumber\\
&&+ \sum_{\alpha\in L,R}\left[v_{\alpha} e^{\frac{i\phi_{\alpha}}{2}}c_{\alpha,1}^{\dagger} d_{\sigma} + H.c.\right].
\label{Ham:transformed}
\end{eqnarray}

Under the above unitary transformation, the Hamiltonian of the central QD, $\mathcal{H}_{\text{QD}}$ remains unaltered, while the SC phase $\phi_{L(R)}$ decouples entirely from the amplitude, $\Delta_0$ (without affecting any of the symmetries of the parent Hamiltonian given in Eq.~\eqref{Eq2}) and appear only as a prefactor, $e^{\frac{i \phi_{L(R)}}{2}}$ in the tunnelling matrix elements. Thus, the decoupling scheme not only simplifies the model but also captures the essential SC features of the leads in the tunnelling part itself, which helps to compute the SC phase difference induced Josephson current for our system. 

For convenience, we define the SC phase difference ($\phi$) between the two KCs as $\phi=\phi_L-\phi_R$. Referring to the variations of static energy spectra presented in Fig.~\ref{fig:static_E_vs_mu} (Appendix~\ref{appendixA}), it is confirmed that the localized MZMs at $E=0$ are indeed present in our static model, which ascertains that the inherent topology of the KC is still intact despite the presence of a QD. This serves as a benchmark for comparison and allows us to contrast our findings with the results reported previously \cite{Taberner2023,Liu2018,Medina-Cuy2023}. 

To compute the static Josephson current ($I(\phi)$) as a function of the SC phase difference, $\phi$, we use the eigenstates ($E_{\gamma}(\phi)$) of $\tilde{\mathcal{H}}_{\text{stat}}$ that satisfy $\tilde{\mathcal{H}}_{\text{stat}}(\phi)\psi_{\gamma}=E_{\gamma}(\phi)\psi_{\gamma}$  for the $\gamma^{th}$ state. Then, the Josephson current can be generically defined in terms of the Free-energy $F(\phi)$ and the Fermi distribution function, $f_{\gamma}=1/2[1-\tanh(E_{\gamma}/2k_BT)]$ as~\cite{Zhang2022, Pal2022, Davydova2022, Cayao2024,Liu2024,Steiner2023}
\begin{equation}
I(\phi) =  \frac{2e}{\hbar}\partial_{\phi} F(\phi),~~~~~~ F(\phi)=\sum_{\gamma}E_{\gamma}(\phi)f_{\gamma}.
\label{def:sum_current}
\end{equation}
For our system, summing over all the filled states below $E=0$ (called as the ground state energy), the above \textit{\text{`sum rule'}} definition of current (\ref{def:sum_current}) at $T=0$ simplifies to
\begin{eqnarray}
I(\phi) = \frac{2e}{\hbar} \frac{\partial}{\partial\phi}E(\phi)= \frac{2e}{\hbar} \frac{\partial}{\partial\phi}\sum_{E_{\gamma}<0}E_{\gamma}(\phi).
\label{def:our_current}
\end{eqnarray}
\par It is important to mention that \( E(\phi) \) satisfies the symmetry relation \( E(\phi) = E(-\phi) \), which stems from the transformation \( \mathcal{U} \tilde{\mathcal{H}}_{\text{stat}}(\phi) \mathcal{U}^{\dagger} = \tilde{\mathcal{H}}_{\text{stat}}(-\phi) \). This implies that both \( \tilde{\mathcal{H}}_{\text{stat}}(\phi) \) and \( \tilde{\mathcal{H}}_{\text{stat}}(-\phi) \) share the same eigenvalues, ensuring that \( E(\phi) \) remains an even function of \( \phi \). Consequently, the Josephson current, \( I(\phi) \), must be an odd function, which satisfies
\begin{equation}
    I(\phi) = -I(-\phi)
    \label{static current_identity},
\end{equation}
which is evidently shown in the current-phase relation of the static Josephson current (Figs.~\ref{fig:static_EI_vs_phi}(a) and (b) in Appendix~\ref{appendixA}). Furthermore, from the static energy and the current variations, it can be understood that in the topological regime ($-2t<\mu<2t$), $E(\phi)$ exhibits a fermion parity switching at $\phi=\pi$ that results in a $4\pi$-periodic Josephson effect with a finite discontinuity in the current profile at $\phi=\pi$~\cite{Kumari2024,Cayao2017}. Whereas, in the trivial limit, these features are absent, ensuring a conventional $2\pi$-periodic Josephson current (for details see Appendix~\ref{appendixA}).

Regardless of weather the system is in the topological or trivial phase, the identity in Eq.\eqref{static current_identity}, reveals that the current is perfectly reciprocal, meaning that a finite diode effect cannot be realized in the static case unless the system is influenced with $\mathcal{TRS}$ and $\mathcal{IS}$ violations. 
For our purposes, as mentioned in Sec.~\ref{intro}, rather than incorporating these additional ingredients, such as an external magnetic field or spin-orbit coupling, we aim to achieve an identical effect by driving the system out of equilibrium. The impact of periodic driving and its role in modifying the transport properties will be explored in detail in the following sections.

\section{Driven scenario}\label{driven}
The junction, as described, does not inherently exhibit any nonreciprocal Josephson current in the static KC-QD-KC heterojunction as the static Hamiltonian lacks explicit violation of $\mathcal{TRS}$ or $\mathcal{IS}$. Typically, realizing a JDE necessitates one or more of the following: an external magnetic field~\cite{Ando2020, Strambini2022} or an intrinsic magnetic moment~\cite{Debnath2024b}, magnetochiral anisotropy~\cite{Baumgartner2022, Bauriedl2022} or a magnetic impurity~\cite{Sun2023, Trahms2023} to be present in the system to achieve a broken $\mathcal{TRS}$. In conjunction to obtain the broken $\mathcal{IS}$, typically the Rashba spin-orbit interaction~\cite{Bauriedl2022, Legg2022, Debnath2024a, Meyer2024}, a band asymmetry~\cite{Soori2023, Hosur2023} or a chiral property of the channel~\cite{Chen2023, Zinkl2022, Nagaosa2023} has to be employed.
However, here we prescribe a much simpler approach where the combined effects of a magnetic field and the spin-orbit interactions can effectively be replicated via (Floquet) driving the system into out-of-equilibrium scenarios, where a carefully chosen periodic modulation can effectively induce a finite nonreciprocity in the Josephson current. 
\subsection{The Floquet formalism}\label{Floquet}
\par We begin by introducing two harmonic drives of the same frequency (however maintaining a finite phase difference) applied to the onsite potentials of the two KCs (as presented in Fig.~\ref{fig:model}), which can be expressed as
\begin{equation}
\mathcal{H}_{F}(t)=\sum_{j=1}^{N_s} [V_1 \cos(\omega t+\zeta) c_{L,j}^{\dagger}  c_{L,j}+ V_2 \cos(\omega t) c_{R,j}^{\dagger}  c_{R,j}].
\label{Drive Hamiltonian}
\end{equation} 
This term, together with Eq.~\eqref{Ham:transformed}, results in a total Hamiltonian for the KC-QD-KC model as
\begin{equation}
\mathcal{H}(t)=\tilde{\mathcal{H}}_{\text{stat}}+\mathcal{H}_{F}(t).
\label{total_driven_Hamiltonian}
\end{equation} 
In Eq.~\eqref{Drive Hamiltonian}, \( V_1 \) and \( V_2 \) represent the driving strengths, which we initially assume to have equal amplitudes, while \( \omega \) denotes the driving frequency and \( \zeta \) captures the phase difference between the two drives. Crucially, the asymmetry introduced by a nonzero \( \zeta \) can break certain symmetries, thereby enabling the emergence of a diode effect in the system, a mechanism that we will explore in detail shortly. Before delving into that, we first provide a pedagogical overview of the Floquet formalism, which will allow us to derive a time-independent effective Hamiltonian for our analysis. 

Floquet theory provides a systematic approach to solve the time-dependent Schrödinger equation by employing the Floquet ansatz, $\ket{\psi(t)} = e^{-i E t} \ket{u(t)}$, where $\ket{u(t+\mathcal{T})}=\ket{u(t)}$ denotes the time-periodic Floquet states, and $E$ represents the Floquet quasienergies. Analogous to quasi-momentum in crystalline solids, these quasienergies are periodic and confined within the Floquet Brillouin Zone, defined as, $E \in [-\pi/\mathcal{T}:\pi/\mathcal{T}]$ ($\mathcal{T} =2\pi/\omega$, period of the drive). On the other hand, the Floquet modes can also be interpreted as the eigenstates of the Floquet stroboscopic time evolution operator ($\hat{U}(\mathcal{T})$) via
\begin{equation}
\begin{split}
    & \hat{U}(\mathcal{T}) \ket{\psi(0)} = \ket{\psi(\mathcal{T})}, \\ & \hat{U}(\mathcal{T}) \ket{u(0)} = e^{-iE\mathcal{T}} \ket{u(\mathcal{T})} = e^{-iE\mathcal{T}} \ket {u(0)},
\end{split}
\end{equation}
with $\hat{U}(\mathcal{T})$ denoted by,
\begin{eqnarray}
    \hat{U}(\mathcal{T})&=&\mathbb{T}  \text{exp} [-\frac{i}{\hbar} \int_{0}^{\mathcal{T}} \mathcal{H} dt] \simeq \exp\left[ -\frac{i}{\hbar} \sum_{i=0}^{N_{\mathcal{T}}-1} \mathcal{H}(\tau_i) \Delta \tau \right]\nonumber\\ 
    &&=\prod_{i=0}^{N_\mathcal{T}-1} \exp\left[ -\frac{i}{\hbar} \mathcal{H}(\tau_i) \Delta \tau \right] + \mathcal{O}(\Delta \tau^2).
\label{evolutionoperator}
\end{eqnarray}
Here, $\mathbb{T}$ denotes the time-ordering product, $\Delta \tau = \mathcal{T}/N_\mathcal{T}$, and $N_\mathcal{T}$ is chosen to be large enough for ensuring convergence. The last expression corresponds to a first-order Trotter decomposition, and the associated error is of the order $\mathcal{O}(\Delta \tau^2)$, which is minimized by choosing sufficiently large time steps, $N_\mathcal{T}$. Thereafter, rewriting Eq. (\ref{evolutionoperator}), one can obtain
\begin{equation}
    \hat{U}(\mathcal{T}) = e^{- i \mathcal{H}_{\text{eff}}\mathcal{T}}.
\end{equation}
Here, $\mathcal{H}_{\text{eff}}$ is the effective time-independent Hamiltonian. Eq.~\eqref{Drive Hamiltonian} fundamentally hosts an eigenvalue equation for the Floquet effective Hamiltonian, $\mathcal{H}_{\text{eff}}$, allowing the quasienergies to be obtained through exact diagonalization of the stroboscopic evolution operator in Eq.~\eqref{evolutionoperator}. This numerically requires a time-ordered decomposition of the evolution operator, which serves as a crucial step in extracting the Floquet effective Hamiltonian, given as
\begin{equation}
\mathcal{H}_{\text{eff}} = \frac{i}{\mathcal{T}} \log \left[ U(\mathcal{T}, 0) \right].
\label{effective_Hamiltonian}
\end{equation}  

This formulation provides a technique to obtain a time-independent Floquet Hamiltonian while ensuring controlled approximation errors. Furthermore, it is essential to recognize that the periodic driving pushes the system into a non-equilibrium state, where the occupation of the quasienergy states deviates from the standard equilibrium distribution.
Throughout our study, we assume a weak coupling between the system and the reservoir, ensuring that the dynamics of the system are predominantly governed by the periodic drive, while the reservoir has minimal influence on the Floquet signatures. In the zero-temperature limit ($T\rightarrow 0$), the time-averaged Floquet Josephson current can be expressed in terms of the quasienergy spectrum ($E_\mathcal{P}$) given as \cite{Kumari2024} (see Appendix~\ref{appendixD} for a detailed derivation)
\begin{equation}
    I= \frac{1}{\mathcal{T}}\int_0^\mathcal{T} I(t^{\prime}) dt^{\prime} = \frac{2e}{\hbar} \sum_{E_{\mathcal{P}}<0} \partial_{\phi} E_{\mathcal{P}}.
    \label{floquet_current}
\end{equation}
\subsection{ Asymmetric Floquet drive: \texorpdfstring{$\mathcal{TRS}$}{TEXT} and \texorpdfstring{$\mathcal{IS}$}{TEXT} breaking}\label{matrix}
breaking
\begin{figure}
\includegraphics[width=0.8\linewidth]{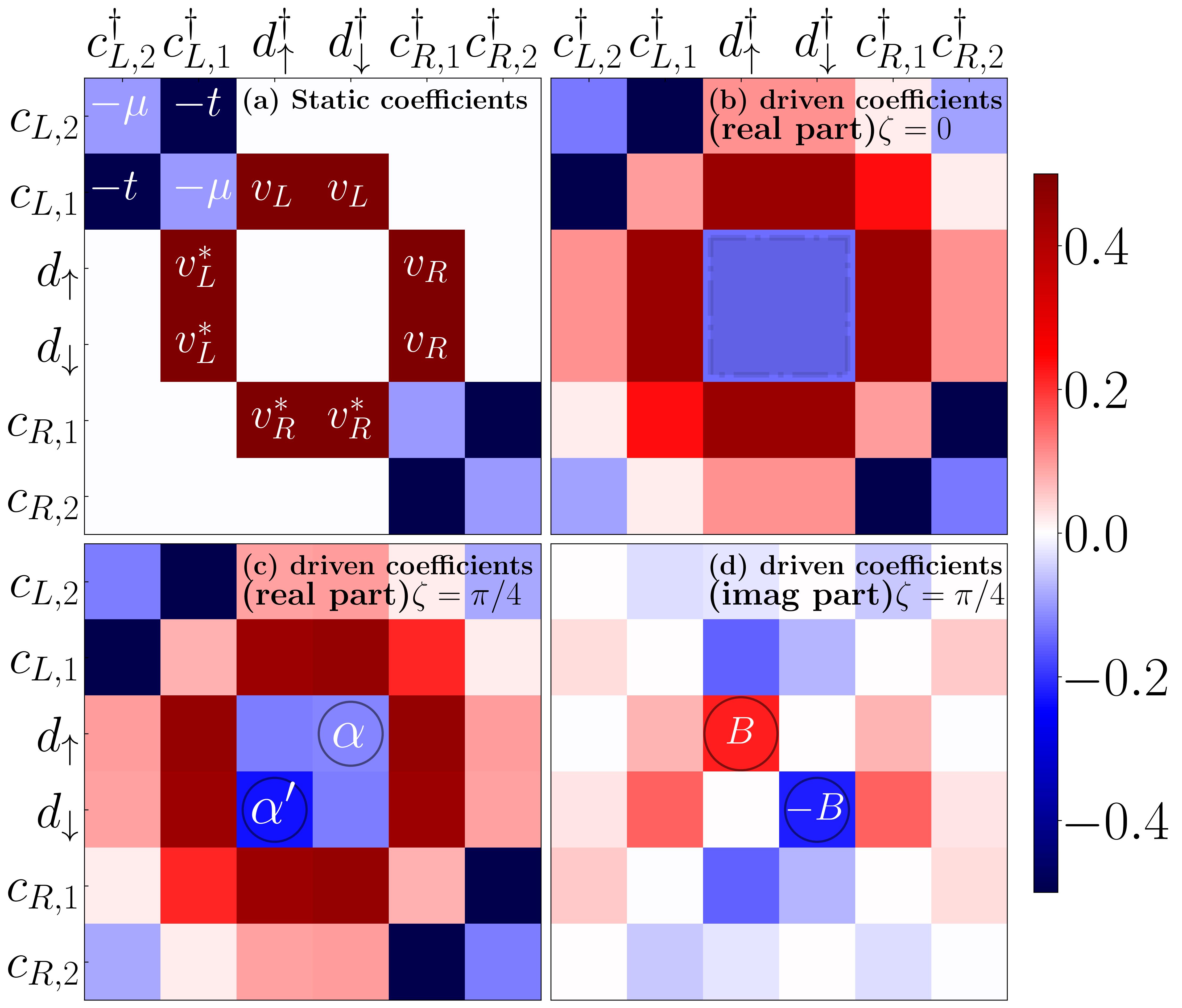}
\caption{The breaking of $\mathcal{IS}$ and $\mathcal{TRS}$ through an asymmetric Floquet drive factor $\zeta$ is illustrated using the Floquet matrix coefficients. While panel (a) displays the static part, panels (b)-(d) refer to the driven version of the model in the coordinate basis ($c_{L,i}^{\dagger}, d_{\sigma}^\dagger, c_{R,i}^{\dagger}$...$c_{L,i}, d_{\sigma}, c_{R,i}$). In the driven case, as long as \( \zeta \neq 0, \pi \), the system induces an artificial spin-orbit coupling \( \alpha (\alpha^{\prime}) \)  and Zeeman-like terms \( B (-B) \), which emerge in the central block as shown in panels (c) and (d) for $\zeta=\pi/4$. This signifies the breaking of \( \mathcal{IS} \) and \( \mathcal{TRS} \) of the central QD. The choice of parameters is indicated by the color bar, with the driving parameters set as \( V_1 = V_2 = 1.5 \) and \( \omega = 3 \). We fix $\epsilon_d=V_g=0$ and $\mu=0.5$.}
\label{fig:matrix}
\end{figure}
Before exploring the characteristics of the Josephson current in the driven system and its role in inducing the nonreciprocal Josephson current by a phase mismatch ($\zeta$) between the two harmonic drives, it is essential to understand how periodic driving disrupts key symmetries by introducing longer-range interactions.
To begin with, in the static case, the $\mathcal{TRS}$ manifests through the relation $\tilde{\mathcal{H}}_{\text{stat}} = \tilde{\mathcal{H}}_{\text{stat}}^*$, while the $\mathcal{IS}$ is characterized by~\cite{Kitaev2001,Leumer2021}
\begin{equation}
    (\mathbb{I}_N \otimes \sigma_z) \tilde{\mathcal{H}}_{\text{stat}}(\Delta) (\mathbb{I}_N \otimes \sigma_z) = - \tilde{\mathcal{H}}_{\text{stat}}(-\Delta)~~\forall \,\, \zeta.
\end{equation}
Here, $\mathbb{I}_N$ is the identity matrix in real space and $\sigma_z$ acts on the particle-hole space. Furthermore, a more comprehensive explanation of the various symmetry operators relevant to the model is provided in Appendix \ref{appendixA}. Now, once the periodic drive is turned on, the $\mathcal{IS}$ is broken, regardless of the value of $\zeta$, that is, 
\begin{equation}
    (\mathbb{I}_N \otimes \sigma_z) \mathcal{H}_{\text{eff}}(\Delta) (\mathbb{I}_N \otimes \sigma_z) \neq - \mathcal{H}_{\text{eff}}(-\Delta).
\end{equation}
On the other hand, $\mathcal{TRS}$ remains intact as long as $\zeta = 0$ or $n\pi$. At values other than these, the relation $\mathcal{H}_{\text{eff}} = \mathcal{H}_{\text{eff}}^*$ no longer holds, signifying the breaking of $\mathcal{TRS}$.

One can alternatively understand the breakdown of these symmetries via considering a Floquet drive, which can manifest artificial magnetic fields or spin-orbit coupling-like terms. To provide a deeper insight, we analyze the matrix configuration corresponding to the Floquet effective Hamiltonian. In Fig.~\ref{fig:matrix}, we numerically compare the matrix representations of the Hamiltonian for both the static and driven versions of the model, with a minimal setup comprising of only two lattice sites. The various elements of the matrix, such as $\mu,t,\Delta_0,v_L,v_R$, etc. are color-coded according to their amplitudes, highlighting the structural modifications introduced by the drive. A key region of interest is the central block (confined within basis $d_{\uparrow},d_{\downarrow}$), which corresponds to the QD. In the static case (Fig.~\ref{fig:matrix}(a)), this block remains empty since the QD energy is set to zero. However, under periodic driving, additional longer-range interaction terms emerge, effectively filling up the central block, as depicted in Fig.~\ref{fig:matrix}(b). Notably, the effective Hamiltonian can host both real and imaginary components, each playing a distinct role in symmetry breaking. Focusing first on the real part, we observe that when the asymmetric phase \(\zeta\) is varied from \(0\) to \(\pi/4\), the two off-diagonal elements acquire different magnitudes ($\alpha$ and $\alpha^{\prime}$), resembling a spin-orbit-like interaction term that breaks $\mathcal{IS}$ (Fig.~\ref{fig:matrix}(c)). Similarly, an inspection of the imaginary components reveals elements that are \textit{not} completely zero, as they should have been in the static scenario, thus signalling a breakdown of $\mathcal{TRS}$. This is further confirmed by the presence of opposite signs for the diagonal elements in the central block, which can be interpreted as an artificial Zeeman field, $B$ (Fig.~\ref{fig:matrix}(d)). Therefore, by tuning only one parameter, that is, \(\zeta\), one can selectively introduce terms in the Hamiltonian that can break both the $\mathcal{IS}$ and $\mathcal{TRS}$, thereby enabling the emergence of diode-like characteristics in the driven system.

\section{Numerical analyses of driven Josephson junction and diode characteristics}\label{numerical}

In this section, we present a comprehensive numerical analysis to explore the key characteristics of the Josephson current, including its current-phase relation (CPR), and the emergence of current nonreciprocity leading to a Josephson diode effect (JDE) under a periodically driven scenario. For such an analysis, the system parameters are set as  $N_s=40$, $\Delta_0 = 0.5t$, $v_L =v_R = 0.5t$, and the driving amplitudes are chosen as $V_1=V_2=1.5t$. The Josephson current is computed in units of $2e/\hbar$. These values have been kept constant throughout the paper. Further, all the energy parameters have been measured in units of the hopping strength $t$, with $t$ being set to unity. In addition, for simplicity, the QD energy ($\epsilon_d$) is kept zero throughout, and also the gate voltage is turned off until its effects are studied in Sec.~\ref{gate tunability}. 
\begin{figure}
\includegraphics[width=0.75\linewidth]{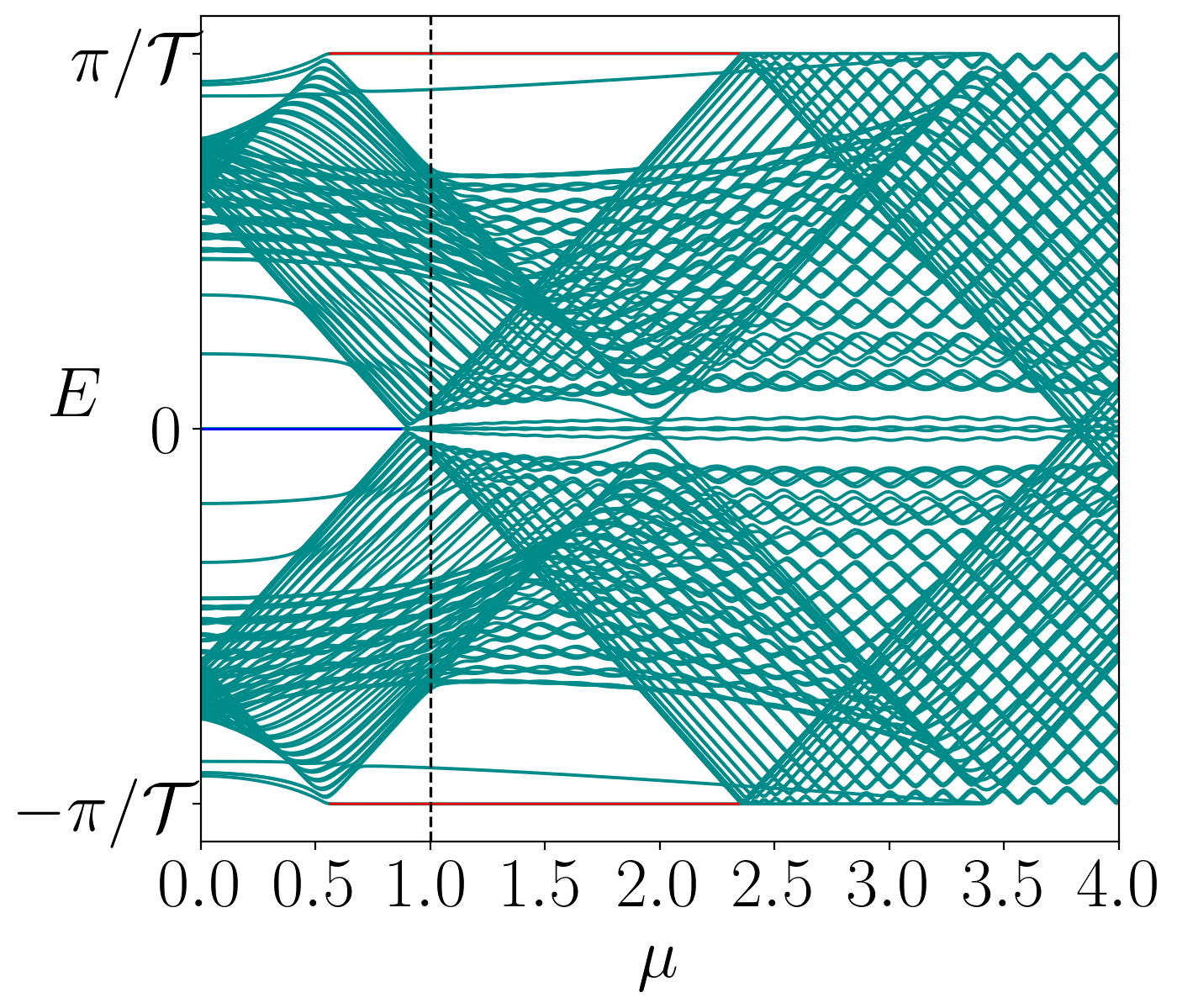}
\caption{Variations of the driven quasienergy spectrum, $E$ (in units of $t$) as a function of the chemical potential of the KC, $\mu$ are shown for the drive amplitudes \( V_1 = V_2 = 1.5 \), the Floquet frequency, \( \omega = 3 \), and asymmetric phase factor, $\zeta=0$. The quasienergies corresponding to the MZM (at $E=0$) and the MPMs (at $E=\pm\pi/T$) are marked in blue and red lines, respectively. This figure vividly illustrates the emergence of both the MZMs, which persist up to \( \mu = 0.9 \), and the MPMs that exist in the range $0.6\leq\mu<2.3$. Other system parameters are set as $N_s=40$, $\Delta_0 = 0.5$, $v_L=v_R=0.5$, and $\epsilon_d=V_g=0$.}
\label{fig:driven_spectrum_vs_mu}
\end{figure}

\subsection{Topological \textit{vs} trivial CPR: Role of Majoranas}\label{driven JC}
To begin with, since a well-defined topological phase is essential for our Josephson junction model to obtain a finite Josephson current, we highlight the inevitable role of chemical potential ($\mu$) in enhancing the topological characteristics of the system in a driven scenario. Fig.~\ref{fig:driven_spectrum_vs_mu} represents the quasienergy spectrum as a function of \( \mu \), with the Floquet frequency ($\omega$) set as \( \omega = 3 \). The figure clearly describes the emergence of both the Majorana zero-modes (MZMs) (which persist up to \( \mu = 0.9 \), marked by the blue line) and the Majorana $\pi$-modes (MPMs) (existing in the range $0.6\leq\mu<2.3$, marked by red lines). Interestingly, unlike hitherto reported findings~\cite{SanJose2014,Cayao2017,Cayao2018} where the Majorana modes localized across the weak link could hybridize at specific values of \( \phi \), our model exhibits no such hybridization, ascertaining the unambiguous existence of the Majorana modes. 
However, it is true that the hybridization of the inner Majoranas (in the vicinity of the QD) depends on the two relative length scales, namely the Majorana localization length ($l_M$) and SC length ($l_S$) of the two KCs~\cite{Cayao2017}. The overlap of two inner Majorana modes being proportional to $e^{-l_S/l_M}=e^{-N_sa/l_M}$~\cite{Gangadharaiah2011}, the hybridization between them is precluded in the limit $l_S\gg l_M$ (as it holds for our case).
This supports the robustness of these Majorana modes, ensuring that the persistence of a finite Josephson current depends entirely on their survival.

\begin{figure}[h]
\includegraphics[width=0.75\linewidth]{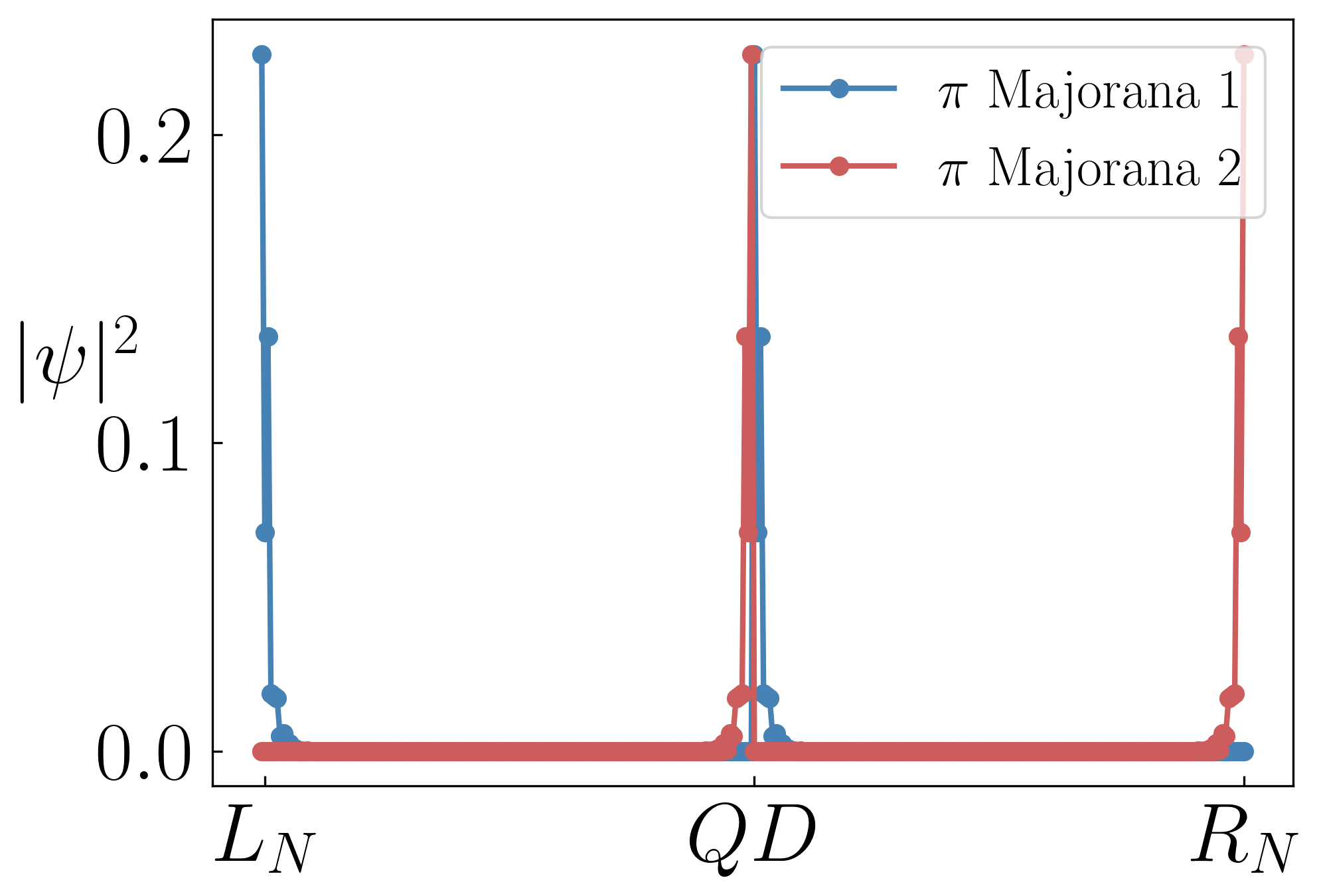}
\caption{The probability distribution profile of the MPMs is presented at \( \mu = 1 \), as the MPMs emerge at $\mu=0.6$ as shown in Fig.~\ref{fig:driven_spectrum_vs_mu}, spanning a system of size \( L = 4N_s + 4 \). The remaining parameters are the same as in Fig. \ref{fig:driven_spectrum_vs_mu}. Regardless of the value of \( \phi \), four Majorana modes are always present, while two of them are localized at the edges of the entire system and the other two are confined at the interfaces of the QD, signifying the topological robustness, despite the presence of an intervening QD between the two $p$-wave KCs.}
\label{fig:driven_probability}
\end{figure}

The presence of such localized edge states can be further validated by examining the probability distribution profile of the Majorana modes across the entire system, which we consider to be of length \( L = 4N_s + 4 \). Fig.~\ref{fig:driven_probability} explicitly depicts the probability distribution of the \( \pi \)-modes at \( \mu = 1 \), revealing the existence of four MPMs (also true for MZMs). Among them, two resemble conventional MPMs, localized at the two ends of the chain, while the remaining two are confined at the interfaces of the QD.

\begin{figure}[t]
\includegraphics[width=\columnwidth]{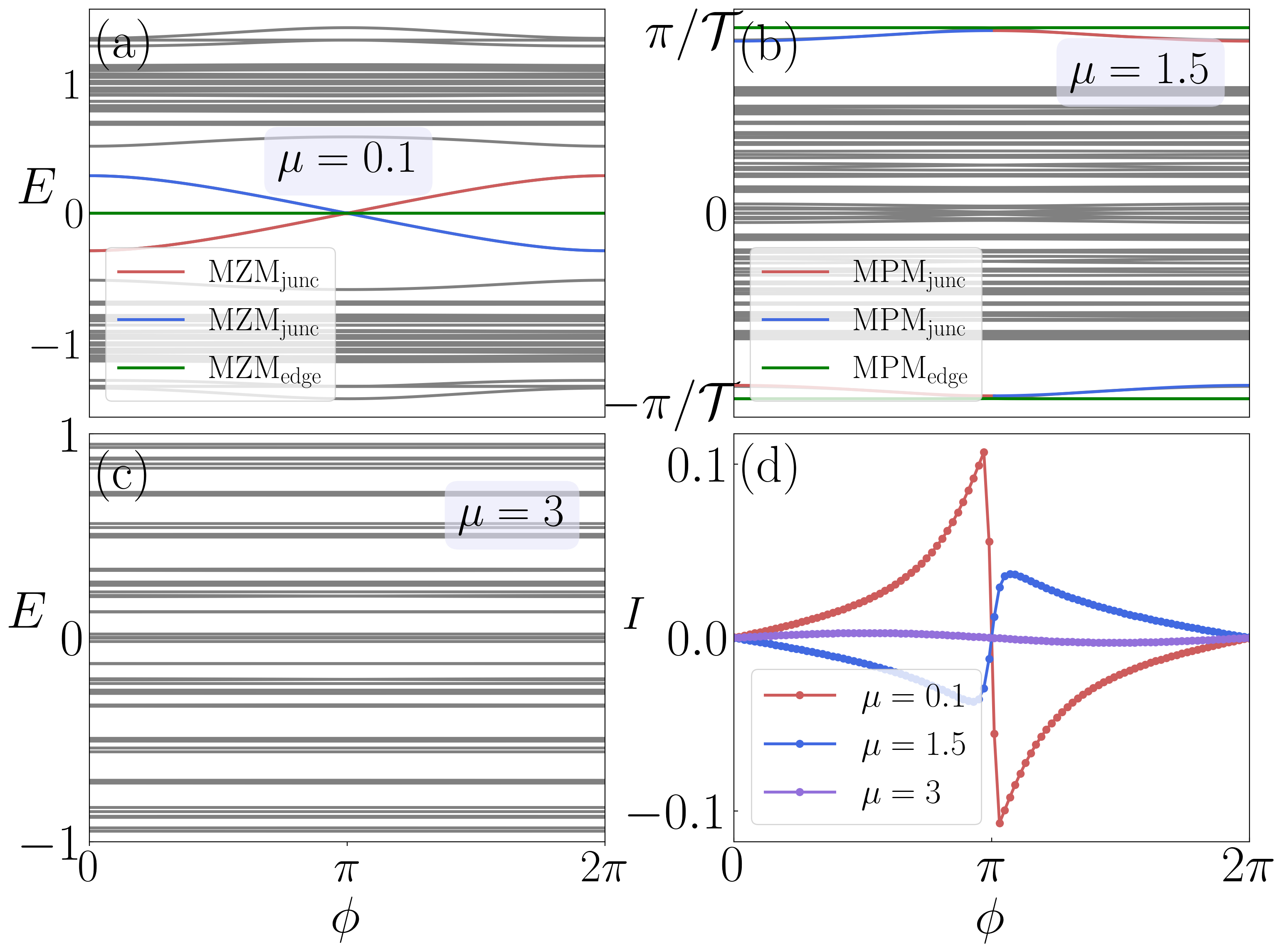}
\caption{Variations of the quasienergy spectra are shown for the topological regimes in (a) and (b) for \(\mu = 0.1\) and \(\mu = 1.5\), respectively, where junction-localized Majoranas (\(\mathrm{MZM}_{\text{junc}}\) and \(\mathrm{MPM}_{\text{junc}}\)) 
become degenerate with their edge counterparts (\(\mathrm{MZM}_{\text{edge}}\) and \(\mathrm{MPM}_{\text{edge}}\))  near \(\phi = \pi\), resulting in a \(4\pi\)-periodic $E$-$\phi$ signature. In contrast, (c) shows the trivial regime (\(\mu = 3\)), where no Majorana modes are present.
(d) illustrates the corresponding CPRs, where for the topological cases they exhibit finite jumps at \(\phi = \pi\), while in the trivial case, a negligibly small bulk-mediated ($2\pi$-periodic) current is seen.
The rest of the parameters are fixed as those in Fig.~\ref{fig:driven_spectrum_vs_mu}.}
\label{fig:current_vs_phi_role_of_topology}
\end{figure}

In accordance with Fig.~\ref{fig:driven_spectrum_vs_mu}, we begin by referring to Fig.~\ref{fig:current_vs_phi_role_of_topology}, which shows the CPR of the periodically driven Josephson current ($I$) for three different values of $\mu$ at a particular drive frequency, namely $\omega=3$. The rationale behind using such a $\omega$ value is understood shortly afterwards. For this plot, we have kept the asymmetric phase factor ($\zeta$) as zero because this analysis only allows us to delineate an optimal range of $\mu$ within which the system remains in the topological regime, facilitating further exploration of its transport characteristics. 
At this moment, it is worth mentioning the role of Majoranas in producing such a finite Josephson current in the topological regime.
Similar to the static case (Appendix~\ref{appendixA}), the $E-\phi$ dispersion gives rise to two outer Majorana modes, denoted as 
$\text{MZM}_\text{edge}$ and 
$\text{MPM}_\text{edge}$, which are highlighted by solid green lines in Figs. \ref{fig:current_vs_phi_role_of_topology}(a) and (b). In addition, we identify two more Majorana modes localized at the junction (labeled $\text{MZM}_\text{junc}$ and 
$\text{MPM}_\text{junc}$) as represented by red and blue solid lines. These junction-localized Majoranas become degenerate with the edge-localized ones specifically at $\phi=\pi$, implying that these junction localized Majorana follows, $E \propto \pm \cos(\phi/2)$, which results in an energy crossing at $\phi=\pi$. Moreover, it indicates a discontinuity as observed in the Josephson current defined as, $I = \sum_{E_p < 0} \partial_\phi E_p \propto \eta \sin(\phi/2)$, where $\eta = \pm 1$ denotes the fermionic parity associated with the occupation of the quasiparticle states. Therefore, the abrupt change in $I$ at $\phi = \pi$ signals a parity switch, which gets manifested as a finite jump in the current (see Fig.~\ref{fig:current_vs_phi_role_of_topology}(d)), a distinctive signature of the $4\pi$-periodic topological Josephson effect~\cite{Kumari2024,Cayao2017}. Interestingly, while the amplitude of the current decreases as $\mu$ approaches the boundary of the topological phase ($0 < \mu < 2.3$), the finite jump at $\phi = \pi$ remains persistent. 
A similar behavior can also be witnessed in the static scenario as described in Fig. \ref{fig:static_EI_vs_phi} of Appendix \ref{appendixA}. 
This emphasizes that, in the topological regime, as long as the junction Majorana modes do not hybridize (i.e., for $l_S \gg l_M$), the Josephson current is entirely Majorana-mediated. However, for \( \mu > 2.3 \), the system enters into a trivial phase, where no such features exist owing to the absence of the Majoranas (as displayed in Fig.~\ref{fig:current_vs_phi_role_of_topology}(c)). The corresponding current profile seemingly appears flat (actually sinusoidal), contributing a negligible bulk current mediated by conventional Cooper pairs similar to the static case. This comparative observation between the bulk- and the Majorana-mediated current fundamentally corroborates that the Josephson current conducts only through the Majorana modes in a KC-QD-KC type Josephson junction, implying the topological dominance of the Majorana-mediated transport.
We present a summary of the above discussions in Table.~\ref{tab:josephson_summary}.

\begin{table*}[t]
\centering
\setlength{\tabcolsep}{10pt} 
\renewcommand{\arraystretch}{1.5} 
\begin{tabular}{l l l l l l}
\hline\hline\\[-1ex] 

\shortstack{\textbf{Current} \\ \textbf{Type}} & 
\shortstack{\textbf{Range} \\ \textbf{of} $\bm\mu$ \textbf{(for} $\bm t\bm=\bm1$}\textbf{)} & 
\shortstack{\textbf{Max} \\ \textbf{Amplitude}} & 
\shortstack{\textbf{Periodicity and} \\ \textbf{Discontinuity}} & 
\shortstack{\textbf{Current-Phase} \\ \textbf{Relation}} & 
\shortstack{\textbf{Ground state} \\ \textbf{Fermion Parity}} \\
\hline\\[-2ex] 

\shortstack{Majorana/\\edge-mediated\\(topological)} & \shortstack{Static: $-2<\mu<2$\\Driven (for $\omega=3$):\\ $0<\mu<2.3$} & $10^{-3}$ - $10^{-1}$ & 
\shortstack{\quad$4\pi$-periodic and \\ jump at $\phi = \pi$} & 
$I \propto \sin(\phi/2)$ & 
\shortstack{Switches at \\ $\phi = \pi$} \\

\\[-1.5ex]
\shortstack{Bulk-mediated\\(trivial)}& \shortstack{Static: $|\mu|\geq 2$\\Driven (for $\omega=3$):\\ $\mu>2.3$} & $10^{-6}$ - $10^{-4}$ & 
\shortstack{$2\pi$-periodic and \\ no jump at $\phi = \pi$} & 
$I \propto \sin(\phi)$ & 
No switching \\

\\[-1.5ex]\hline\hline

\end{tabular}
\caption{Distinguishing characteristics between Majorana- and bulk-mediated Josephson currents, which assert that the maximum current contribution is mediated through topological modes (either MZMs or MPMs), exhibiting $4\pi$-periodic CPR, whereas the trivial bulk regime contributes only weak, and conventional $2\pi$-periodic supercurrents.}
\label{tab:josephson_summary}
\end{table*}

\begin{figure}
\includegraphics[width=1\linewidth]{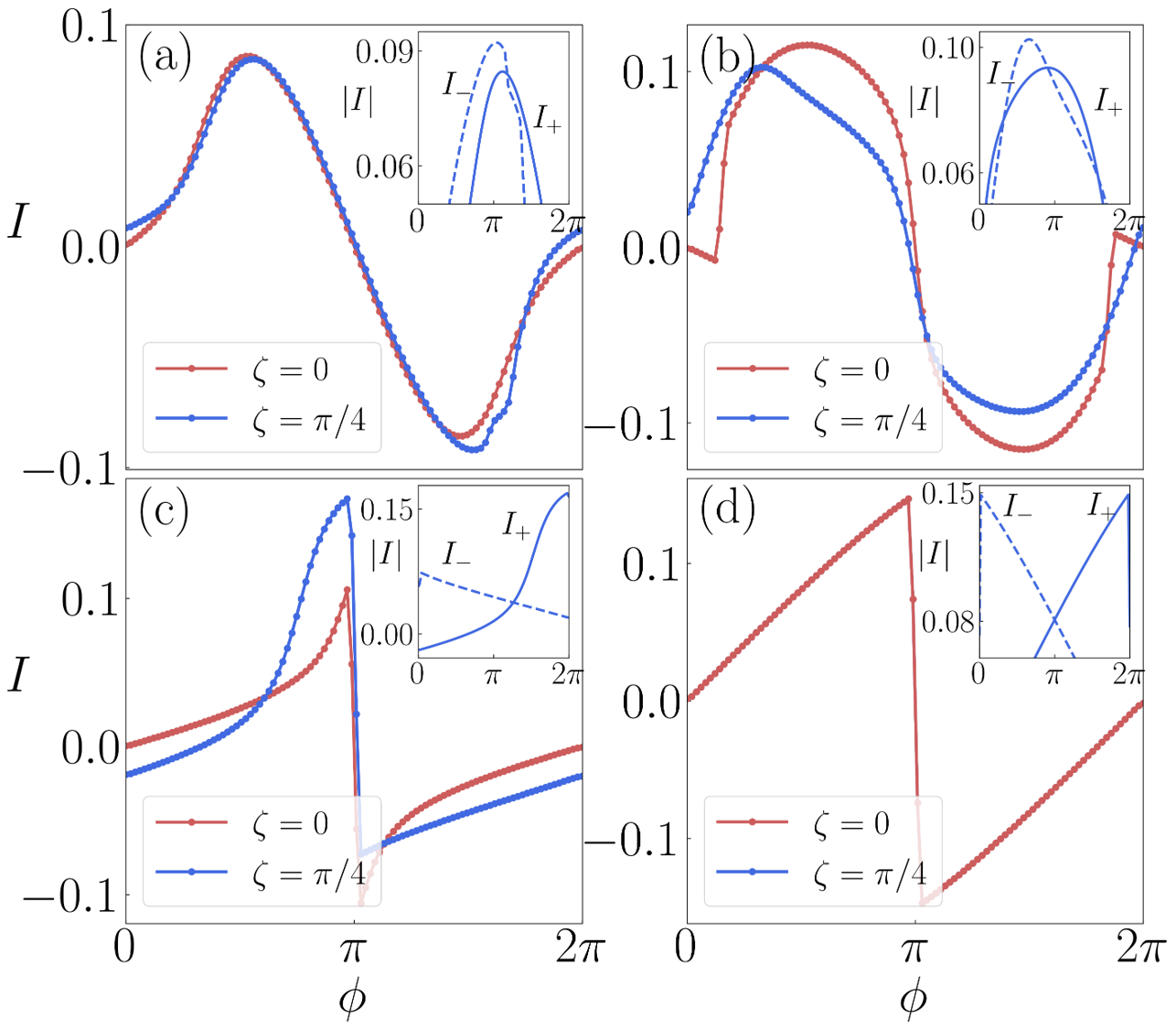}
\caption{Time-averaged Josephson current, $I$ (in units of $2e/\hbar$) vs. the SC phase difference $\phi$ is plotted at $\mu=0.5$ for $\zeta=0$ and $\zeta=\pi/4$ labeled by red and blue lines, respectively and shown in lower to higher Floquet frequency ($\omega$) regimes as: (a) $\omega=0.8$, (b) $\omega=2$, (c)~$\omega=3$, and (d)~$\omega\geq 5$. The insets (plotted for $|I|$ of the blue curves at $\zeta=\pi/4$) represent the disparity between two differently directed currents, namely $I^+$ and $I^-$ designated by solid and dotted blue lines, respectively. Insets of Figs.~(a)-(c) signify prominent disparity between $I^+$ and $I^-$ in the lower to intermediate range of $\omega$, which ceases to exist in Fig.(d) at high $\omega$ reaching the static limit. Other system parameters are fixed as those in Fig.~\ref{fig:driven_spectrum_vs_mu}. 
}
\label{fig:current_vs_phi}
\end{figure}
\subsection{Nonreciprocity in the CPR and its tunability}\label{driven JC2}
To illustrate how different driving parameters, such as the drive frequency ($\omega$) and the asymmetric phase ($\zeta$) influence the CPR of the driven current ($I$), we present the results in Fig.~\ref{fig:current_vs_phi}. 
We first examine the case where \(\zeta = 0\). It is expected that in the high-frequency regime, the system behaves similarly to its static counterpart (Appendix~\ref{appendixA}), resulting in a CPR characterized by a $E$-$\phi$ relation that supports an energy level crossing, which in turn leads to an abrupt `jump' at \(\phi = \pi\) (Fig.~\ref{fig:current_vs_phi}(d)). A similar behavior is observed at certain intermediate frequencies, such as \(\omega = 3\) (see Fig.~\ref{fig:current_vs_phi}(c)). In contrast, for lower frequencies, such as \(\omega = 0.8\) (Fig.~\ref{fig:current_vs_phi}(a)) and \(\omega = 2\) (Fig.~\ref{fig:current_vs_phi}(b)), the Josephson current deviates significantly from the static case, and instead exhibits a nearly sinusoidal profile. This deviation arises due to the absence of a uniform bulk gap at low driving frequencies. A more detailed analysis of the current behavior in this regime is provided in Appendix \ref{appendixE}. Notably, in the low-frequency regime, the lack of a well-defined bulk gap compromises the protection of Majorana modes, leading the system toward a trivial phase, where the current exhibits a smooth $2\pi$ periodicity and is predominantly carried by Cooper pairs. 

A more fundamental and prime aspect of our investigation is to achieve the nonreciprocity of Josephson current, which leads to a finite diode rectification for which the breaking of both $\mathcal{TRS}$ and $\mathcal{IS}$ are mandatory. It is discussed elaborately in Sec.~\ref{matrix} that introducing an additional phase difference between the two drives, namely, \(\zeta\) can effectively generate an artificial magnetic field and a spin-orbit-like terms in the Hamiltonian (as presented in Fig.~\ref{fig:matrix}) which leads to the breaking of these symmetries. Hence, we study the effect of $\zeta$, specifically for \(\zeta = \pi/4\) (the choice for such $\zeta$-value will be clear as we proceed) on CPR of the driven Josephson current. Together, the artificial Zeeman and spin-orbit terms cause an imbalance in the dynamics of Majorana quasiparticles moving in opposite directions, that is, a forward-moving pair can experience a different effective velocity with respect to the oppositely moving pairs.
This directional dependency of current leads to an inequality in the critical currents~\cite{Davydova2022, Pal2022}, expressed as \( I_c^+ \neq I_c^- \), where \( I_c^+ = \max(I(\phi)) \) and \( I_c^- = \min(I(\phi)) \), with \(+\) and \(-\) signs denoting the direction of the current. Consequently, the CPR becomes asymmetric satisfying
\begin{equation}
    I(\phi) \neq -I(-\phi), 
    \label{ajc}
\end{equation} 
which is evident in Fig.~\ref{fig:current_vs_phi} and the plots are denoted by the blue lines. A closer examination of Fig.~\ref{fig:current_vs_phi} (insets of Fig.~\ref{fig:current_vs_phi}(a)-(c)) further highlights the relative difference between \( I_c^+ \) and \( I_c^- \), which plays a crucial role in quantifying the nonreciprocity in terms of the diode RF (defined in Eq.~\eqref{diode_RF}). This is the central result of our study. However, before concluding the findings of Fig.~\ref{fig:current_vs_phi} and analyzing the diode characteristics explicitly by estimating the RF, we comment on a few more crucial attributes of the driven CPR for completeness.

Although the driven Josephson current deviates from a purely sinusoidal form in both low and intermediate frequency regimes (Fig.~\ref{fig:current_vs_phi}(a)-(c)), yet it gives rise to an anomalous $\phi_0$-Josephson current~\cite{Mayer2020, Davydova2022, Martin2009, Yokoyama2014, Assouline2019,Taberner2023,Soori2023b}, that is $I(\phi)$ attains a nonzero value at $\phi=0,\pi$ whenever the asymmetric phase factor is nonzero.
As it is known that $\phi_0$-Josephson junction is likely to be realized in presence of the Zeeman and spin-orbit coupling terms~\cite{Mayer2020, Davydova2022, Martin2009, Yokoyama2014, Assouline2019}, the plausible occurrence of anomalous $\phi_0$-Josephson current (especially for $\omega<5$) characteristics in our study is therefore mediated through a nonzero $\zeta$. This is also a direct consequence of the broken $\mathcal{TRS}$ and $\mathcal{IS}$ phenomenon caused by $\zeta$ presented in Fig.~\ref{fig:matrix}. 
\begin{figure}
\includegraphics[width=0.75\linewidth]{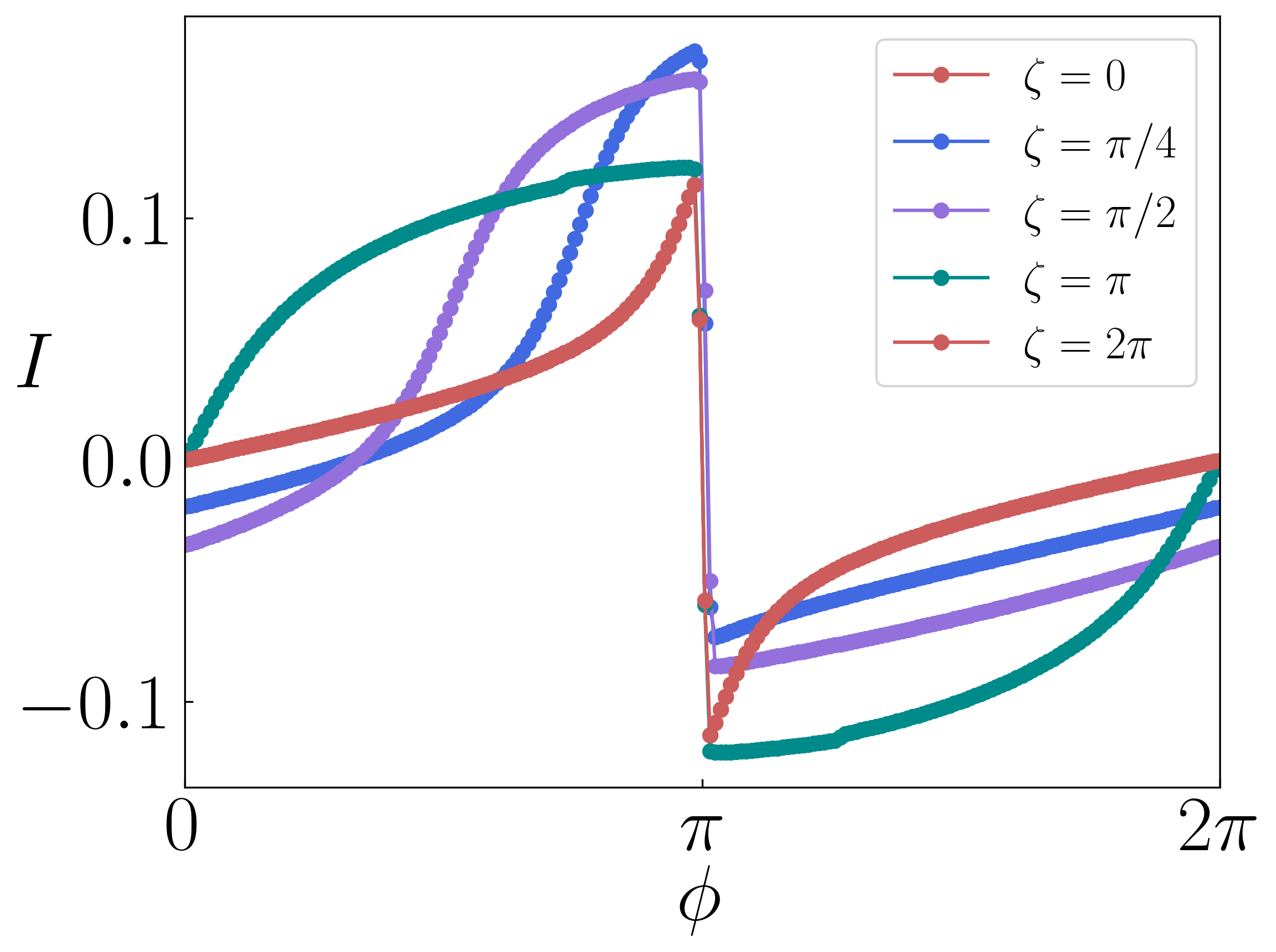}
\caption{The driven CPR in terms of the time-averaged Josephson current, $I$ (in units of $2e/\hbar$) vs. the SC phase difference $\phi$ is plotted at a particular Floquet frequency, $\omega=3$ for different values of the asymmetric Floquet drive parameter, $\zeta$. Other system parameters are fixed as those in Fig.~\ref{fig:driven_spectrum_vs_mu}. The current does not render any nonreciprocity when $\zeta=n\pi$ ($n=0,\pm1,\pm2,..$). However, for intermediate values of $\zeta$, such as $\zeta=\pi/4$, it shows a finite nonreciprocity.}
\label{fig:current_vs_phi_for_diff_zeta}
\end{figure}
\begin{figure}
\includegraphics[width=0.75\linewidth]{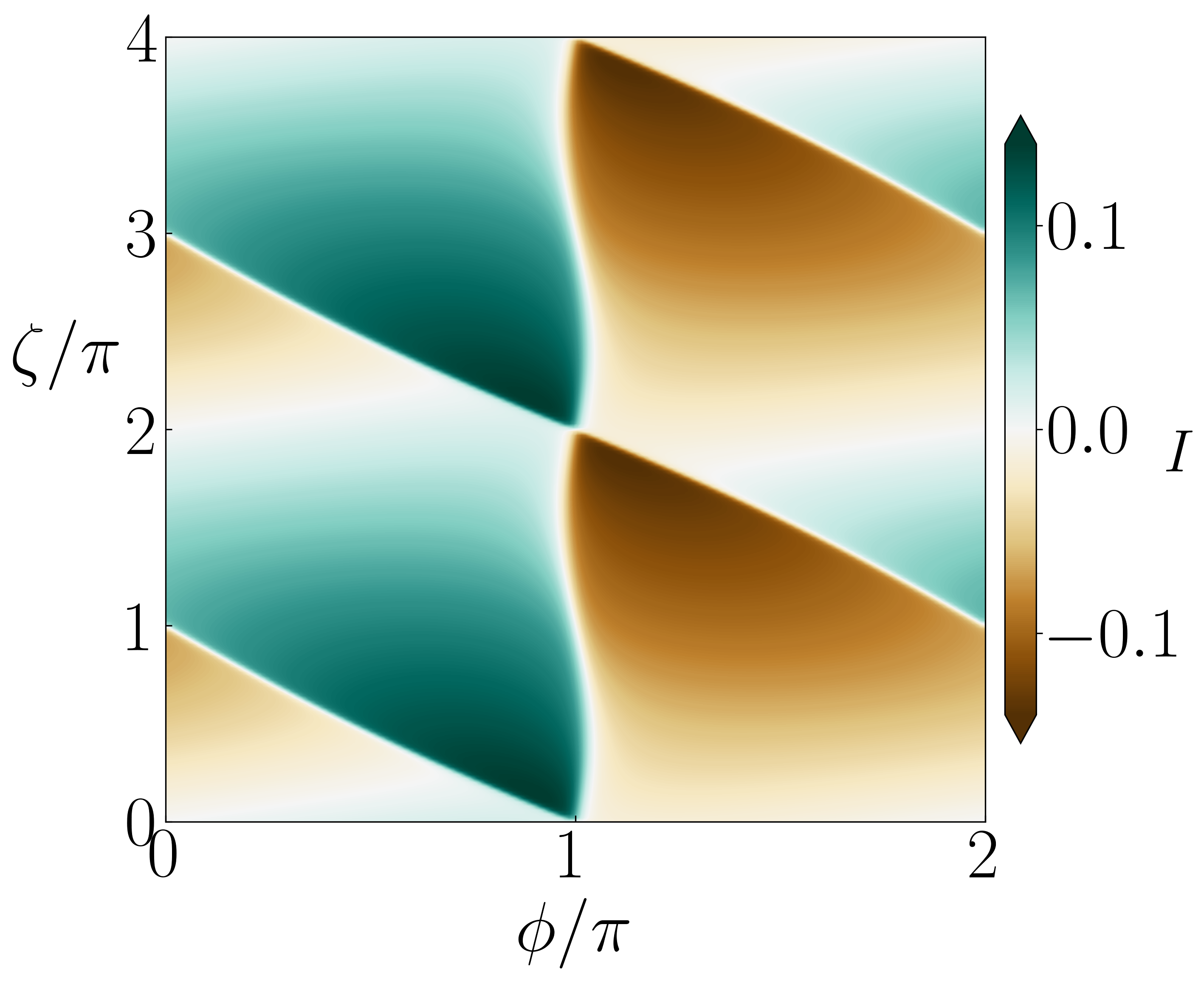}
\caption{A contour plot of the time-averaged Josephson current, $I$ (in units of $2e/\hbar$) is shown in the $\zeta-\phi$ plane (scaled by $\pi$) for a particular Floquet frequency, $\omega=3$. Other system parameters are fixed as those in Fig.~\ref{fig:driven_spectrum_vs_mu}. For specific values of $\zeta$, $I$ remains purely non-sinusoidal, supporting the nonreciprocity. Hence, the tunability of the nonreciprocity is controlled by the drive parameters.}
\label{fig:current_vs_zeta_phi}
\end{figure}
Now, we present Fig.~\ref{fig:current_vs_phi_for_diff_zeta} at a particular frequency, namely, $\omega=3$ that displays how the driven Josephson current varies for different values of the asymmetric phase factor \(\zeta\), which demonstrates tunability of the nonreciprocal current. Clearly, the current satisfies $I_c^+=I_c^-$ \textit{iff} $\zeta=\pm n\pi$, where $n=0,1,2, ...$, corroborating the vanishing of the nonreciprocity. Furthermore, as the current profiles for \(\zeta = 0\) and \(\zeta = 2\pi\) become equal, it indicates a \(2\pi\)-periodicity of the Josephson current with respect to \(\zeta\). In contrast, for intermediate values such as \(\zeta = \pi/4\) and \(\pi/2\), the current largely deviates from the sinusoidal behavior (maintaining the discontinuity as it shown in Fig.~\ref{fig:current_vs_phi} for an intermediate regime of $\omega$) and more importantly there exists a significant distinction between $I_c^+$ and $I_c^-$,  which ensures the existence of nonreciprocity in the driven picture for these values of $\zeta$. Because of this reason, we (mostly) fix $\zeta$ at $\zeta=\pi/4$ whenever needed for our numerical analysis.  Further, to gain a more comprehensive understanding, we plot the phase diagram of the Josephson current in the \(\zeta\)-\(\phi\) plane in Fig.~\ref{fig:current_vs_zeta_phi}. The results confirm that for specific values of \(\zeta\) (that are relevant for the nonreciprocity), the Josephson current remains purely non-sinusoidal, yet periodic with respect to both $\phi$ and $\zeta$. Moreover, extending the analysis to negative values of \(\zeta\) reveals an important symmetry relation of the time-averaged current $I$ with respect to $\phi$ and $\zeta$ as
\begin{equation}
   I(\phi, \zeta) = -I(-\phi, -\zeta), ~~\forall~~\zeta\neq 0.
   \label{I_phi_zeta_identity}
\end{equation}
\subsection{Diode efficiency: Rectification factor}\label{RF}
By this time, we have established the notion of a finite nonreciprocity of Josephson current (solely induced by the asymmetric phase factor, $\zeta$ of the Floquet drive) in our system, which is further responsible for the JDE. Now, in this section, we explicitly estimate how \textit{efficient} such a diode can be owing to this nonreciprocity.
Recalling the fact that for any value of \(\zeta\) (excluding \(\zeta = \pm n\pi\)), there exists a disparity between \( I_c^+ \) and \( I_c^- \), which fundamentally ascertains the emergence of the JDE. This observation naturally motivates us to compute the diode RF, $\mathcal{R} =\left(I_c^+ - |I_c^-|\right)/\left(I_c^+ + |I_c^-|\right)$ , following Eq.~\ref{diode_RF}.

\begin{figure}
\includegraphics[width=0.75\linewidth]{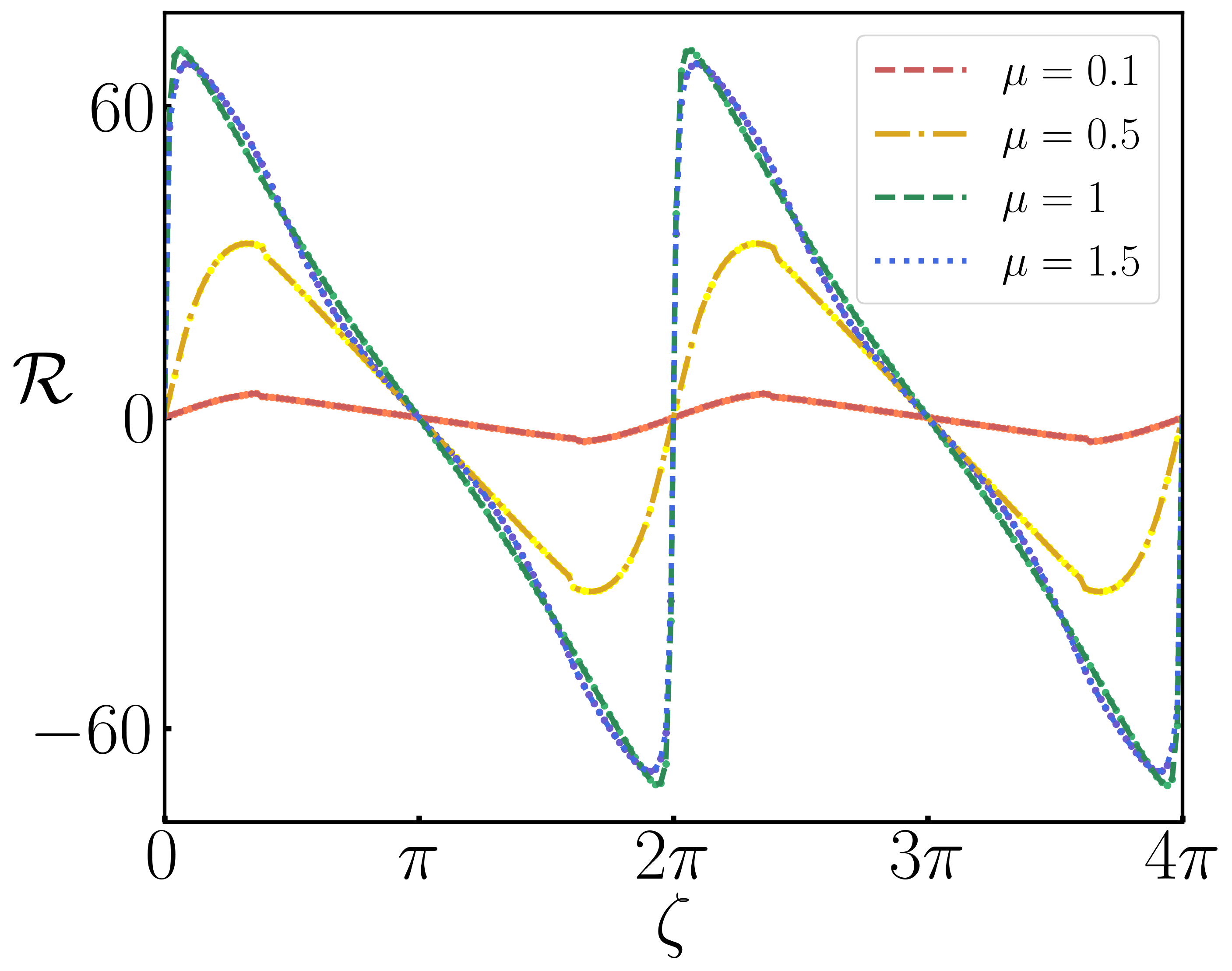}
\caption{Variations of the diode RF ($\mathcal{R}$) as a function of the asymmetric Floquet drive parameter, $\zeta$ are displayed for different values of the chemical potential ($\mu$) of the KCs at a chosen Floquet frequency, $\omega=3$. Other system parameters are fixed as those in Fig.~\ref{fig:driven_spectrum_vs_mu}. It is shown that $\mathcal{R}$ exhibits a sawtooth variation, and it is $2\pi$-periodic with respect to $\zeta$. Furthermore, $\mathcal{R}$ undergoes a sign reversal at every $\pi$ interval of $\zeta$, which can also be understood from Fig.~\ref{fig:current_vs_zeta_phi}. The maximum $\mathcal{R}$ achieved for this range of $\mu$ is around $70\%$ at specific values of $\zeta$.}
\label{fig:RC_vs_zeta}
\end{figure}
As $\zeta$ is the main driving agent for generating a finite RF, we present Fig.~\ref{fig:RC_vs_zeta} to investigate how \(\mathcal{R}\) varies as a function of \(\zeta\) for different values of the chemical potential \(\mu\) at a fixed Floquet frequency, $\omega$. However, we keep the QD parameters fixed as $\epsilon_d=V_g=0$. The dependence of $\mathcal{R}$ on $\mu$ points to a sawtooth-like variation as a function of $\zeta$ regardless of the choice of \(\mu\). However, the magnitude of \(\mathcal{R}\) increases with increasing \(\mu\), yet its oscillatory behavior with respect to \(\zeta\) remains consistent and reaches a maximum of $\mathcal{R}$.  
Hence, our inspection reveals that the maximum $\mathcal{R}$, namely $\mathcal{R}_{\text{max}}$ can be achieved around $\mathcal{R}_{\text{max}}\sim 70\%$ for our KC-QD-KC Josephson junction at $\zeta=(2n\pi\pm\pi/8)$ provided $\mu$ and the driving parameters ($\zeta$ and $\omega$) are preferably chosen. However, it is important to note that the presence of a finite $V_g$ can significantly vary $\mathcal{R}$ (both qualitatively and quantitatively) for these values of $\mu$.

Another key feature is that the $2\pi$-periodic \(\mathcal{R}\) (with respect to $\zeta$) changes its sign at every \(\pi\) interval of \(\zeta\). This sign reversal of \(\mathcal{R}\) can be further verified through the contour plot of \(I\) in the \(\phi\)-\(\zeta\) plane (Fig.~\ref{fig:current_vs_zeta_phi}), which clearly shows that at each \(\pi\)-interval of \(\zeta\), the Josephson current acquires an additional \(\pi\) phase. This leads to a crucial relation in the RF satisfying
\begin{equation}
    \mathcal{R} (\zeta) = - \mathcal{R} (\zeta + \pi).
    \label{RF_identity}
\end{equation}
Thus, we conclude that \(\zeta\) not only plays a key role in producing a finite JDE, but also acts as a tuning parameter (responsible for the sign reversal of $\mathcal{R}$) for setting the biasing condition, offering an effective means of controlling the JDE.

\begin{figure}
\includegraphics[width=\linewidth]{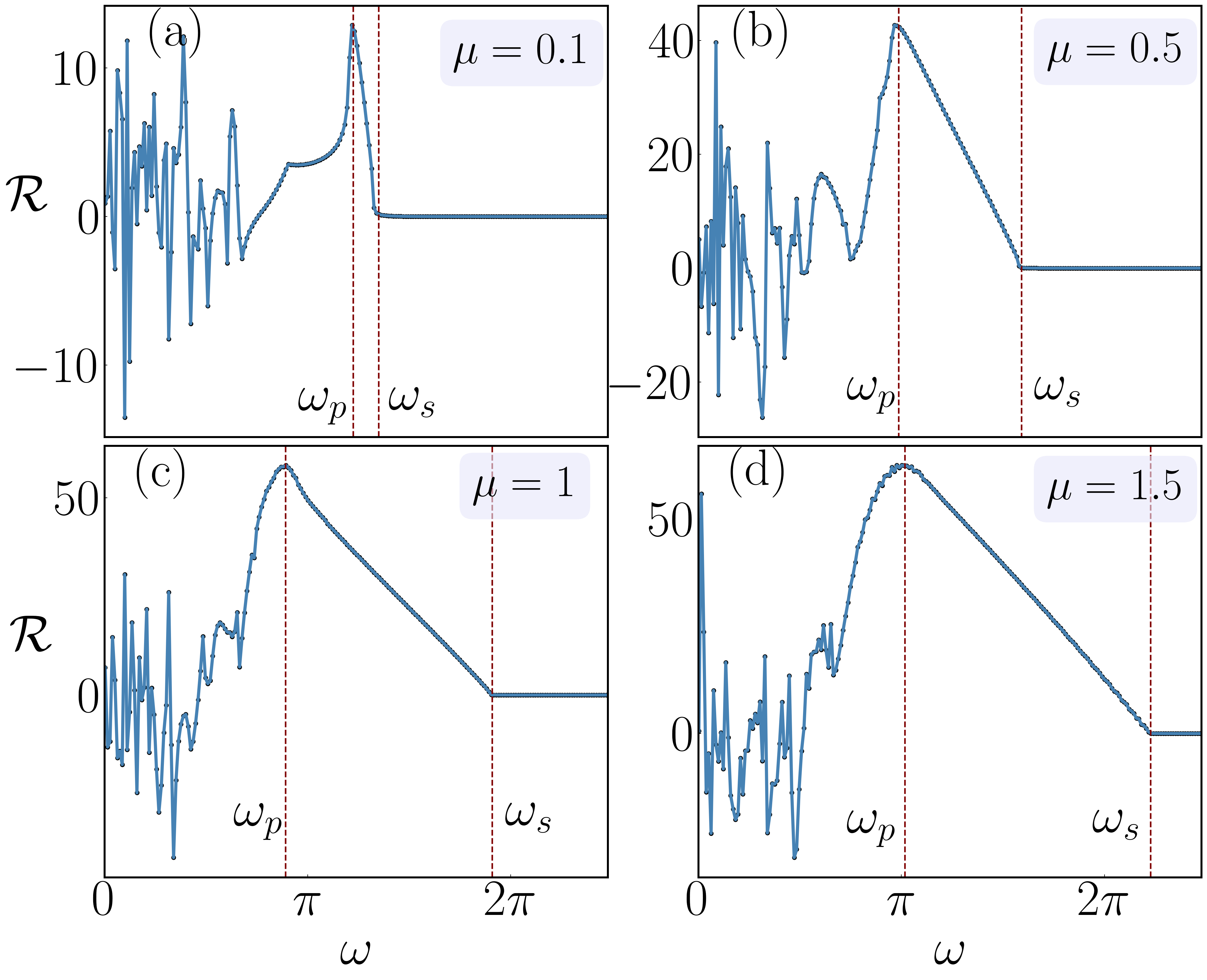}
\caption{Dependence of the RF ($\mathcal{R}$) on the Floquet frequency, $\omega$ is displayed for different values of the chemical potential ($\mu$) as (a)~$\mu=0.1$, (b)~$\mu=0.5$, (c)~$\mu=1$, (d)~$\mu=1.5$. Here, $\zeta$ is fixed at $\pi/4$. Other system parameters are fixed as those in Fig.~\ref{fig:driven_spectrum_vs_mu}. These variations reveal that for each case of $\mu$, $\mathcal{R}$ exhibits a peak at a certain $\omega=\omega_p$ accompanied by oscillations in the lower $\omega$ range, and then declines linearly in the intermediate $\omega$ range and finally becomes zero at $\omega=\omega_s$ where it reaches the static limit. The rate of decline is different for different values of $\mu$.}
\label{fig:RC_vs_omega}
\end{figure}

To gain deeper insight into how the diode effect operates across different frequency regimes, it is essential to analyze the RF as a function of Floquet frequency, \(\omega\). Fig.~\ref{fig:RC_vs_omega} presents the variation of \(\mathcal{R}\) with \(\omega\) for four distinct values of $\mu$. As said earlier, the asymmetric drive parameter, $\zeta$ is fixed at $\pi/4$ for our analysis. A noteworthy observation from this figure is that regardless of the choice of \(\mu\), \(\mathcal{R}\) exhibits irregular fluctuations in the low-frequency regime. This irregularity stems from the absence of a uniform pattern or well-defined periodicity in the Josephson current in the low-frequency regime, a feature that we have explained elaborately in Appendix \ref{appendixE}. On the other hand, as $\omega$ increases \(\mathcal{R}\) reaches a peak for each of these values of $\mu$ (exhibiting a maximum of $\mathcal{R}$ around $70\%$ for $\mu=1.5$) at a certain $\omega$ ($\omega_p$), before undergoing a linear decline in the intermediate-frequency regime (the frequency range where \(\mu\) is comparable to \(\omega\)). This linear dependence indicates that within this regime of $\omega$, the Josephson current should exhibit a smooth and periodic variation. Moreover, in a similar fashion, the physical origin behind such a linear dependency has been illustrated in Figs.~\ref{fig:current_at_different_frequencies} and \ref{fig:gap_invariant} of Appendix \ref{appendixE}. Consequently, RF gradually decreases and eventually vanishes in the high-frequency regime, which is expected in the static limit. Hence, this notable feature implies that the transition from a finite to a zero value of $\mathcal{R}$ occurs at a critical $\omega$ (say $\omega_s$) which shifts to a greater value for larger values of \(\mu\). Moreover, as \(\mu\) increases, the rate of decline in rectification in the intermediate-frequency regime gets slower, exhibiting a larger difference between $\omega_p$ and $\omega_s$. Therefore, for larger values of $\mu$, $\omega_s$ is required to be larger for the RF to vanish. This analysis demonstrates a high-precision smooth operating region for the diode by controlling the frequency $\omega$ and the asymmetric drive strength $\zeta$.

In summary, by appropriately tuning both \(\zeta\) and \(\omega\), one can effectively control and optimize the RF based on the specific requirements. Moreover, the ability to modulate rectification through external driving parameters could be instrumental in designing high-efficiency SC circuits, nonreciprocal quantum devices, and next-generation SC electronics.

\subsection{Gate tunability of the diode effect}\label{gate tunability}
So far, we have discussed the effects of an asymmetric drive in elucidating the occurrence of a finite rectification for such a Majorana coupled Josephson diode. It is also crucial to understand the importance of a QD playing the role of a weak link in this context. As pointed out earlier, the discrete energy level of the QD ($\epsilon_d$) can be adjusted through electrostatic gating~\cite{Alivisatos1996,Zwerver2022,Burkard2023} which offers a precise control on Josephson current~\cite{Gupta2023, Taberner2023, Yan2025, Meyer2024, Trahms2023, Cheng2023, Debnath2024a}, enabling a fine-tuning of the critical current and the phase dynamics. Therefore, it is essential to examine how the QD energy level (tuned by a gate voltage \( V_g \)) can impact both the Josephson current and the RF. 
The application of the gate electrode effectively modifies $\epsilon_d$ to $\epsilon_d^\prime=\epsilon_d-eV_g$, which further suggests that $eV_g$ can act as a rescaled QD energy. That is true even if $\epsilon_d=0$ as kept throughout this study. Another important aspect of utilizing the QD as the weak link is its ability to act as a discrete resonant level and to cause successive on and off resonances with respect to the Fermi-energy (chemical potential, $\mu$) of the KCs (SC leads)~\cite{Dam2006, Herrero2006}. This significantly influences the periodic modulation of the Josephson current and hence the RF. Thus, this control of level-matching feature through tuning $V_g$ is the key to manifest transport aided by the Majoranas across such QD-based Josephson junction. 
Therefore, tuning $V_g$ becomes necessary to exploit the role of the QD on the Josephson current and hence the diode's RF.
\begin{figure}
\includegraphics[width=\linewidth]{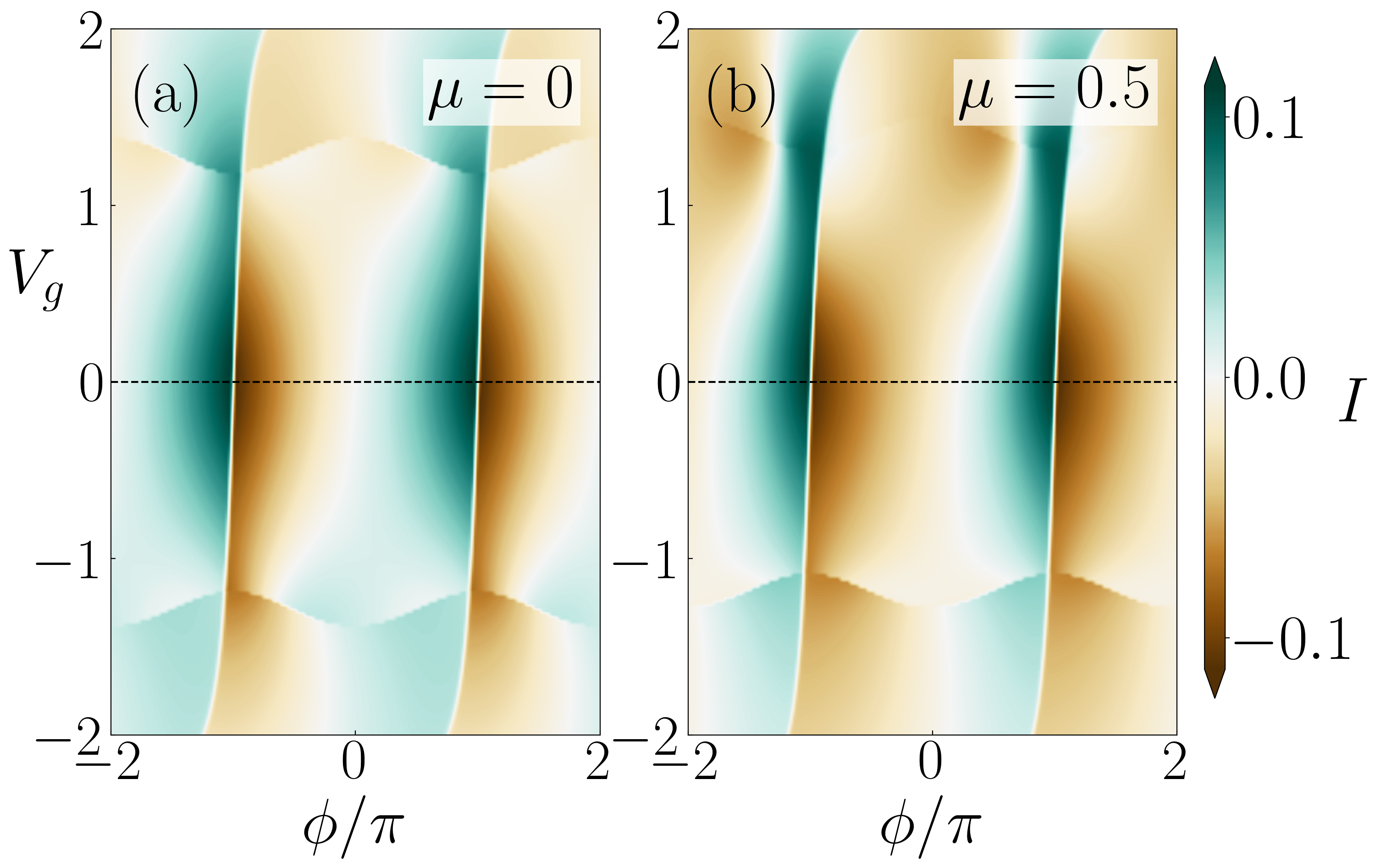}
\caption{The combined effects of an external gate voltage ($V_g$) and the chemical potential ($\mu$) on the time-averaged Josephson current, $I$ is displayed in a contour plot, where $I$ (in units of $2e/\hbar$) is plotted in the $V_g-\phi$ plane for a particular Floquet frequency, $\omega=3$, and asymmetry phase factor $\zeta = \pi/4$. Panel (a) shows the scenario for $\mu=0$, while panel (b) shows that for $\mu = 0.5$. Other system parameters are set as those in Fig.~\ref{fig:driven_spectrum_vs_mu}.}
\label{fig:Josephson_current_vs_phi_vs_Vg}
\end{figure}
\begin{figure}
\includegraphics[width=0.75\linewidth]{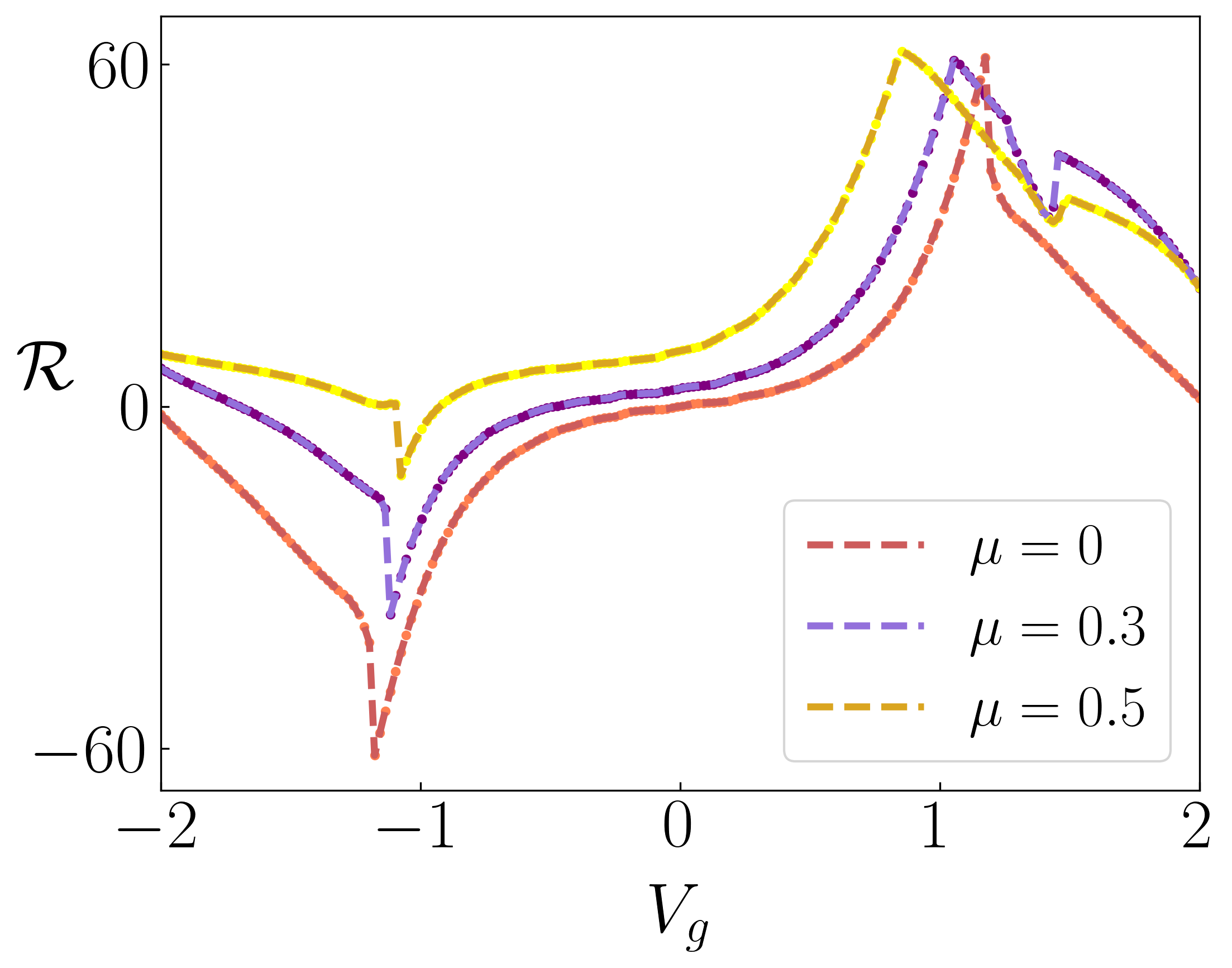}
\caption{The figure illustrates the variation of $\mathcal{R}$ as a function of an external gate voltage, $V_g$, for different values of the chemical potential, $\mu$ at a particular Floquet frequency, $\omega=3$, and asymmetry phase factor $\zeta = \pi/4$. The rest of the system parameters are chosen as mentioned in Fig.~\ref{fig:driven_spectrum_vs_mu}. Clearly, $\mathcal{R}$ has a symmetric pattern with respect to $V_g=0$ only when $\mu=0$. Moreover, $\mathcal{R}$ increases with increasing $V_g$. The fine adjustment of $\mu$ and $V_g$ determines the tunability of the diode and hence the optimization of $\mathcal{R}$.}
\label{fig:RC_vs_Vg}
\end{figure}

To begin with, we explore the CPR as a function of \( V_g \). For the analyses presented in Fig.~\ref{fig:Josephson_current_vs_phi_vs_Vg}, the parameters are chosen exactly the same as those kept in Sec.~\ref{driven JC2}. At this stage, it is essential for us to recall that to sustain a finite Josephson current, \( eV_g \) must lie within the SC gap, $2\Delta_0$~\cite{Dam2006, Herrero2006}, i.e., \( V_g \in [-1,1] \) corresponding to the SC gap chosen as $\Delta_0=0.5$. However, this condition does not appear to be entirely true in the driven scenario since in a Floquet system, the SC gap is no longer fixed at $\Delta_0=0.5$, instead evolves to an effective gap, namely $\Delta_{0}^{\text{eff}}$, whose magnitude depends on both the strength and frequency of the drive. Therefore, for a given choice of \( \Delta_0 = 0.5 \), we fix the range of $V_g$ as \( -2 \leq V_g \leq 2 \) accordingly for generating the contour plot for $I$ in the \( V_g\)-\( \phi \) plane in Fig.~\ref{fig:Josephson_current_vs_phi_vs_Vg} and also for the variations of $\mathcal{R}$ for different values of $\mu$. A careful examination of Fig.~\ref{fig:Josephson_current_vs_phi_vs_Vg}(a) reveals a key identity persistent in the time-averaged current, given by 
\begin{equation}
    I(\phi, V_g) = -I(-\phi, -V_g)~~~\text{iff}~~~\mu=0.
    \label{I_phi_Vg_identity}
\end{equation}
As Eq.~\eqref{I_phi_Vg_identity} suggests, this symmetry with respect to \( V_g = 0 \) (as shown in Fig.~\ref{fig:Josephson_current_vs_phi_vs_Vg}(a)) holds only when the chemical potential ($\mu$) is identically equal to zero. Any deviation from \( \mu = 0 \) causes a disparity in the current profile (higher contribution along positive $V_g$ axis for $\mu>0$) around $V_g=0$, thereby rendering Eq.~\eqref{I_phi_Vg_identity} as invalid at those values of $\mu$ which can be understood from Fig.~\ref{fig:Josephson_current_vs_phi_vs_Vg}(b).

Further, by taking this into account, we analyze the variation of \( \mathcal{R} \) as a function of \( V_g \) for different values of \( \mu\) at $\zeta=\pi/4$ which is illustrated in Fig.~\ref{fig:RC_vs_Vg}. As anticipated, much like the Josephson current, \( \mathcal{R} \) retains its symmetry around \( V_g = 0 \) for $\mu=0$ and adheres to the relation: $\mathcal{R}(V_g) = -\mathcal{R}(-V_g)$.
Interestingly, akin to the behaviour of $I$, the symmetry of \( \mathcal{R} \) around \( V_g = 0 \) also gets disrupted (accompanied by an upward shift in \( \mathcal{R} \)) accordingly when \( \mu \) deviates from zero (shown in violet and yellow dotted lines). Our findings indicate that a finite rectification effect persists within an extended window of \( V_g \in [-2:2] \) in the driven case. Furthermore, within this range $\mu$, the maximum rectification \( \mathcal{R}_{\max} \approx 70\% \), can also be achieved, albeit at a finite $V_g$. However, $\mathcal{R}_{\text{max}}$ can be enhanced further by increasing $\mu$ or by choosing a different set of parameters (not shown here). 
The enhancement of the RF with $|V_g|$ for $\mu\neq 0$ indicates a better degree of resonance between the Fermi energy of the KC leads ($\mu$) and the QD energy ($\sim -eV_g$). However, beyond a certain $|V_g|$, the barrier potential across the QD caused by this level mismatch precludes the flow of Josephson current, rendering a decrease in $I$ and hence in the RF. Thus, the interplay between $\mu$ and $V_g$ strongly influences the current dynamics and the diode efficiency.
In summary, both the Josephson current and \( \mathcal{R} \) exhibit a periodic dependence on \( V_g \) (provided \( \mu = 0 \)), and we have successfully identified an optimal range of \( V_g \) where a substantial and hence technologically significant (tunable) diode effect can be realized.
\section{Summary and conclusion}\label{summary}
To summarize, we investigate the JDE under the application of an asymmetric Floquet drive to the two 1D $p$-wave KCs (acting as two SC leads) and a QD (playing the role of a weak link) sandwiched between them, mimicking a prototype for a KC-QD-KC type Josephson junction. We began by introducing the static version of our model and briefly discussed the essential topological physics of localized MZMs in the static spectrum. Thereafter, we show that the static Josephson junction lacks nonreciprocity due to the absence of an essential symmetry breaking phenomenon and the CPR satisfies $I(\phi) = -I(-\phi)$. However, when we apply a periodic Floquet drive to the KCs with different phases (with the asymmetric phase factor $\zeta$), $\mathcal{TRS}$ and $\mathcal{IS}$ are broken simultaneously. 
Hence, we numerically calculate the behaviour of the time-averaged Floquet Josephson current ($I$) and elaborately investigate its CPR, varying our system's parameters. 
At first, we decipher a preferred range of $\mu$ to access the essential topological effects of the Majoranas in asserting a $4\pi$-periodic CPR (because of fermion parity switching), while a negligible bulk-mediated sinusoidal current in the trivial regime is observed. It is worth mentioning that beyond theoretical predictions, signatures of the $4\pi$-Josephson effect have also been reported in various experiments~\cite{Wiedenmann2016,Laroche2019,Yu2018}.

The central outcome of our study is realized as soon as we turn on the asymmetric phase factor, $\zeta$ which notably distinguishes the disparity between the forward ($I^+$) and reverse ($I^-$) Josephson current (satisfying ($I^+\neq I^-$)) owing to the simultaneous breaking of $\mathcal{TRS}$ and $\mathcal{IS}$, which leads to a finite nonreciprocity and rectification. Additionally, the application of $\zeta$ brings in other important attributes of the Josephson junction phenomena, such as a \text{``anomalous Josephson junction CPR''} or $\phi_0$-Josephson current satisfying $I(\phi=0,\pi)\neq 0$ and a non-sinusoidal pattern of Josephson current, relevant for attaining a finite nonreciprocity for our diode setup. 
Next, we quantify the efficiency of our KC-QD-KC diode for various system parameters. The explicit variation of rectification factor ($\mathcal{R}$) as a function of $\zeta$ reveals its $2\pi$-periodic sawtooth behaviour along with a sign change characteristic at every $\pi$-interval of the phase $\zeta$ satisfying $\mathcal{R}(\zeta)=-\mathcal{R}(\zeta+\pi)$. Our results conclude that the chemical potential of the KCs and the driving parameters play a crucial role in tuning and optimizing the largest possible rectification for such a diode. Gate tunability of the QD energy level additionally modulates the rectification. By judiciously choosing the system parameters, the maximum rectification can be achieved around $\mathcal{R}\sim70\%$ for our diode setup.

From experimental perspectives, QDs can be fabricated using semiconductors~\cite{Alivisatos1996,Zwerver2022,Burkard2023}, such as GaAs, InSb, etc., or a single \text{`adatom'} (metallic or non-metallic nature) acting as a Josephson junction weak link. By controlling the radius of the QD (under strong confinement effects), a single energy level can be attained which is further tunable by electrostatic gating~\cite{Alivisatos1996,Zwerver2022,Burkard2023,Dam2006,Herrero2006}. 
Although the experimental propositions on Majorana have been speculated in the past~\cite{Mourik2012, Stevan2014}, engineering of these modes remains elusive till now~\cite{brouwer2021, Zhang2021}. However, recent works on the realization of a minimal KC using an array of QDs~\cite{Dvir2023, Samuelson2024} have been put forward to validate the existence of the topological Majorana modes.
An experimental setup for our model can be conceptualized using an Nb sheet~\cite{Deng2012} or a ceramic superconductor, such as ${\text{YBa}_{2}\text{Cu}_{3}\text{O}_{7}}$ (YBCO) as the bulk with magnetic atoms or QDs with magnetic impurities being deposited via an STM-probe~\cite{Eigler1990} on it like an array of adatoms that induces Majorana modes at the ends of the KCs~\cite{Kim2018, Schneider2023}.  Applying two different ac voltage drives~\cite{Gabelli2013} to the KCs or using a photonic simulator~\cite{Cheng2019}, one can generate an asymmetric Floquet drive, by which modulation of the chemical potential of the KCs can be achieved. Finally, connecting a dc source to the intermediate QD will facilitate tuning the gate voltage~\cite{Gupta2023,Yan2025,Alivisatos1996,Zwerver2022,Burkard2023}.

To conclude, though attempts have been made to establish a diode effect either induced by different magnetic and spin-orbit coupling effects or by a Floquet drive (to the intermediate weak link), our model not only assists in engineering a nonreciprocal supercurrent associated with topological Majorana modes, but also presents an alternative to such external agencies. Therefore, our study underscores a distinct proposition of a QD-based Floquet Josephson diode, replacing the conventional usage of such external effects via driven KCs, which will aid in realizing Majorana-mediated quantum transport in SC devices and quantum computation.
\section*{Acknowledgments}
We sincerely thank Prof. Diptiman Sen for his valuable comments and suggestions. D.D. acknowledges the Department of Space (DoS), Government of India, for all the support at the Physical Research Laboratory and the Department of Physics, IIT Guwahati, for the local hospitality
during her visit to pursue this work. D.D. also thanks P. Dutta for her encouragement and support. K.B. acknowledges the Science and Engineering Research Board (SERB), Govt. of India, for providing financial support through the National Post Doctoral Fellowship (NPDF) (File No. PDF/2023/000161). 
\appendix
\section{Topological \textit{vs} trivial characteristics of the static energy and current }\label{appendixA}
The MZMs, being one of the fundamental characteristics of a KC, can be visualized in the energy spectrum of the entire system, which is obtained by diagonalizing the transformed Hamiltonian (\ref{Ham:transformed}). 

\begin{figure}
\begin{center}
\includegraphics[width=0.9\linewidth]{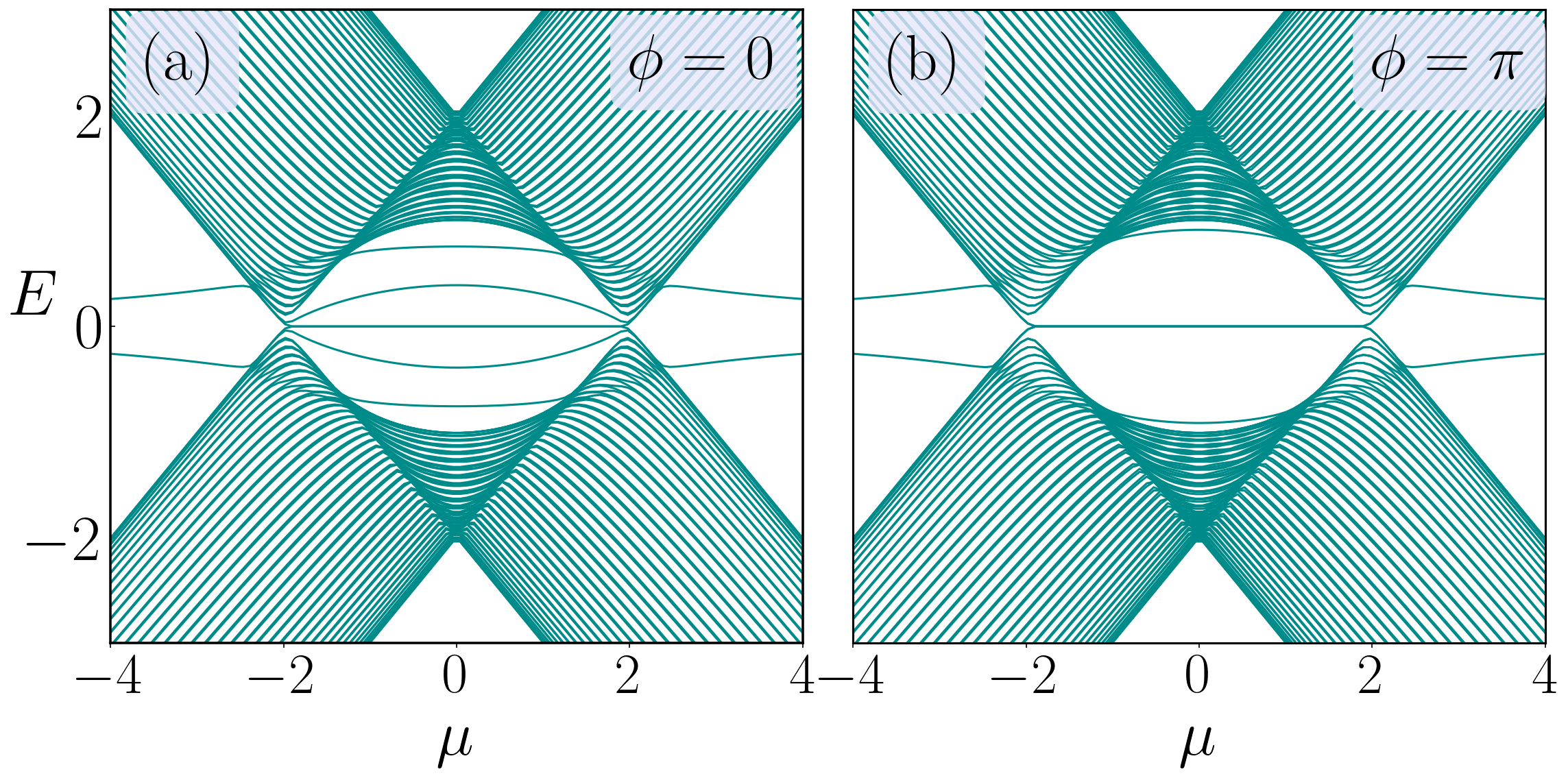}
\caption{Variation of the static energy spectrum, $E$ (in unit of $t$) with respect to the chemical potential of the KC, $\mu$ is shown for (a) $\phi=0$ and (b) $\phi=\pi$. The system parameters are set as $N_s=40$, $\Delta_0 = 0.5$, $\epsilon_d = V_g=0$, and $v_L=v_R=0.5$. 
At $\phi=\pi$, the system comprises of both edge and junction localized Majoranas that appear at the same energy ($E=0$). On the other hand, at $\phi=0$, the junction localized Majoranas deviate from the edge Majoranas and appear as in-gap states as illustrated in panel (a).}  
\label{fig:static_E_vs_mu}
\end{center}
\end{figure}

For $\phi_{L(R)}=0$ and $\pi$, $\mathcal{H}_{\text{L(R)}}$ preserves both the $\mathcal{TRS}$, namely, $\hat{\mathcal{K}} \hat{\mathcal{H}}_{L(R)}(k) \hat{\mathcal{K}} = \hat{\mathcal{H}}_{L(R)}(-k)$ ($\hat{\mathcal{K}}$ being an anti-unitary operator that typically acts as complex conjugation ($\mathcal{K}$) in the Majorana basis). Further, the particle-hole symmetry (PHS) given by, $\hat{\sigma}_x\hat{\mathcal{K}} \hat{\mathcal{H}}_{L(R)}(k) \hat{\sigma}_x\hat{\mathcal{K}} = -\hat{\mathcal{H}}_{L(R)}(-k)$ is obeyed as well. These symmetries refer for the \text{``BDI''} topological class, while for $\phi_{L(R)}\neq0,\pi$ it only obeys PHS which belongs to the topological \text{``D''} class~\cite{Ryu2010,Tong2013}. Also, $\mathcal{H}_{L(R)}$ exhibits other symmetries, namely the inversion symmetry denoted via $\hat{\sigma}_z \hat{\mathcal{H}}_{L(R)}(k,\Delta) \hat{\sigma}_z = \hat{\mathcal{H}}_{L(R)}(k,-\Delta)$, and the chiral symmetr, namely, $\hat{\sigma}_x \hat{\mathcal{H}}_{L(R)}(k) \hat{\sigma}_x = -\hat{\mathcal{H}}_{L(R)}(k)$. 
It can be seen from the Fig.~\,\ref{fig:static_E_vs_mu} that both $\mathcal{TRS}$ and PHS are preserved both for $\phi=0$ and $\phi=\pi$, which manifest in the appearance of two outer MZMs (localized at the ends of the chain). However, for $\phi=0$, there exist in-gap states coming from the inner MZMs (localized at the junction of the QD) in Fig.~\ref{fig:static_E_vs_mu}(a) which become degenerate with outer MZMs at $E=0$ for $\phi=\pi$ (Fig.~\ref{fig:static_E_vs_mu}(b)). Thus, this figure demonstrates that despite the presence of a weak link (such as a QD) in between two KCs, the MZMs (also the MPMs for the driven case) indeed exist as localized modes.
\begin{figure}
\includegraphics[width=\columnwidth]{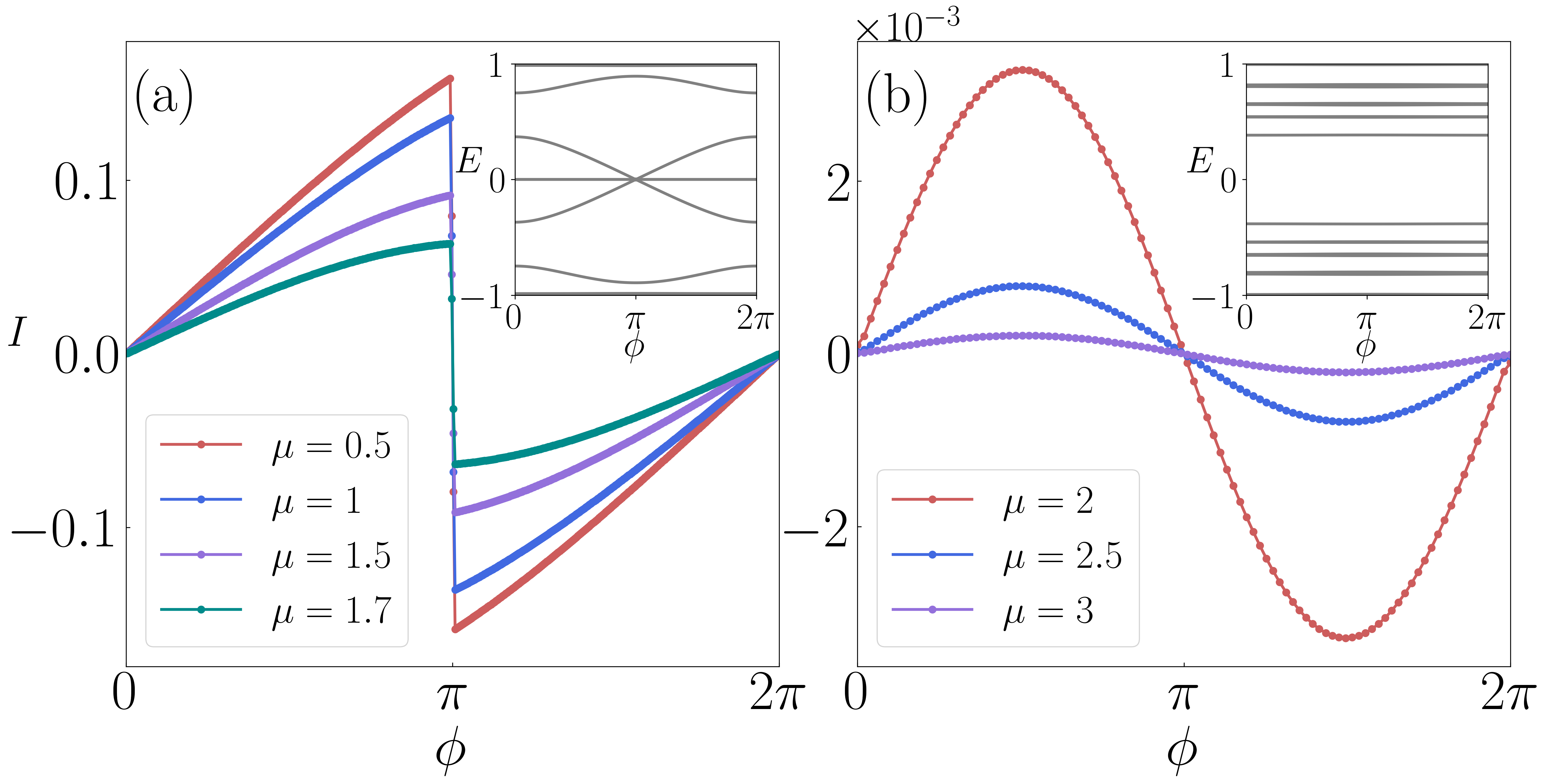}
\caption{CPR corresponds to the topological regime ($|\mu|<2t$) is shown in (a), indicating a $4\pi$-periodicity with a finite jump at $\phi=\pi$ in the (Majorana-mediated) current profile. The inset represents the $E-\phi$ dispersion for $\mu=0.5$. In contrast, (b) illustrates the CPR at the trivial regime ($|\mu|>2t$), where the Majoranas are absent, indicating a smooth $2\pi$-periodic sinusoidal bulk-mediated current. The rest of the parameters are fixed as those in Fig.~\ref{fig:static_E_vs_mu}.}
\label{fig:static_EI_vs_phi}
\end{figure}

It is noteworthy that the in-gap states observed in Fig. \ref{fig:static_E_vs_mu}(a) correspond to junction-localized Majorana modes, which are responsible for the finite jump observed in the Josephson current owing to the energy level crossing (parity switching) at $\phi = \pi$ (inset of Fig. \ref{fig:static_EI_vs_phi}(a)). Fig. \ref{fig:static_EI_vs_phi}(a) presents examples of such Majorana-mediated $4\pi$-periodic Josephson junction signatures for various values of $\mu$ within the topological regime. Interestingly, although the amplitude of the current decreases with increasing $\mu$, the finite jump at $\phi = \pi$ remains robust in all cases. In contrast, Fig. \ref{fig:static_EI_vs_phi}(b) depicts the Josephson current in the trivial regime, where the absence of Majorana modes results in a smooth and sinusoidal current predominantly carried by Cooper pairs.
\section{Derivation of the Floquet Josephson current}\label{appendixD}
Here we present a detailed derivation of the Floquet Josephson current formula. 
Following the approach of Ref.~\cite{Kumari2024}, to analyze the current-phase relation, it is essential to recognize that the Floquet drive forces the system out of equilibrium, leading to a non-thermal occupation of the quasienergy states. To account for this, we model the system as being weakly coupled to an external thermal reservoir with coupling strength $V^{\nu}$. Assuming the reservoir has a flat (energy-independent) density of states, the time-dependent Schrödinger equation governing the system can be written as
 \begin{equation}
i\hbar\frac{\partial}{\partial t} |\Psi(t)\rangle = [\mathcal{H}(t) - i \Gamma^{\nu}] |\Psi(t)\rangle. \label{EOM}
\end{equation}
Here, $\Gamma^{\nu} = V^{\nu} \rho^{\nu} V^{\nu \dagger}$ describes the density of states ($\rho^{\nu}$) of the reservoirs and their coupling ($V^{\nu}$) with the system. Given the periodic nature of the driving, characterized by $\mathcal{H}(t+\mathcal{T})= \mathcal{H}(t)$, where $\mathcal{T}$ is the drive period, and the solution of Eq.~(\ref{EOM}) takes the form, $|\Psi_{\mathcal{P}}(t) \rangle = e^{\left(-\frac{i E_{\mathcal{P}}}{\hbar} - \gamma_{\mathcal{P}}\right)t}|u_{\mathcal{P}}(t)\rangle$.
The states $|u_\mathcal{P}(t)\rangle$ are time-periodic Floquet modes, which can be expanded by a Fourier series as
$|u_\mathcal{P}(t)\rangle = \sum_m |u_\mathcal{P}^{m}\rangle e^{-i m \omega t},$ where $\omega = \frac{2\pi}{\mathcal{T}}$ is the driving frequency. Accordingly, the steady-state density matrix $\hat{\rho}(t)$ can be expressed in the Floquet mode basis, $|u_\mathcal{P}(t)\rangle$, of the isolated system as
\begin{eqnarray}
\hat{\rho}(t) = \frac{1}{\mathcal{T}} \int_0^\mathcal{T} dt \sum_{\mathcal{P},\mathcal{Q}} \hat{n}_{\mathcal{P}\mathcal{Q}}(t) |u_\mathcal{P}(t)\rangle \langle u_\mathcal{Q}(t)|.
\end{eqnarray}
The current operator can be written in terms of the
steady-state density matrix, $\hat{\rho}(t)$ and the Floquet states $|u_\mathcal{P}(t)\rangle$ as 
\begin{eqnarray}
\hat{I}(t) &=& \text{Tr} (\hat{\rho}(t) \partial_{\phi}\hat{\mathcal{H}}(t))
    = \sum_{\mathcal{P}} \langle u_{\mathcal{P}}(t)|\hat{\rho}(t) \partial_{\phi}\hat{\mathcal{H}}(t)|u_{\mathcal{P}}(t)\rangle \nonumber\\
&=&\sum_{\mathcal{P} \mathcal{Q}}  \langle u_{\mathcal{P}}(t)| \hat{\rho}(t) | u_{\mathcal{Q}}(t)\rangle \langle  u_{\mathcal{Q}}(t)| \partial_{\phi}\hat{\mathcal{H}}(t) |u_{\mathcal{P}}(t)\rangle.\nonumber\\
\label{current} 
\end{eqnarray}
To calculate $\langle u_{\mathcal{P}}(t)| \rho(t) | u_{\mathcal{Q}}(t)\rangle\equiv \hat{n}_{\mathcal{P}\mathcal{Q}}(t)$, we decompose the thermal average of the operator $\hat{n}_{\mathcal{P}\mathcal{Q}}(t)$ into its Fourier components as, $n_{\mathcal{P}\mathcal{Q}}(t) = \sum_{\lambda} e^{-i \lambda\omega t} n_{\mathcal{P}\mathcal{Q}}^{\lambda},$ where the Fourier coefficients $ n_{\mathcal{P}\mathcal{Q}}^{\lambda}$ are given by~\cite{Kumari2024}
\begin{equation}
n_{\mathcal{P}\mathcal{Q}}^{\lambda} = \sum_{\nu, m} \int_{-\infty}^{\infty}\frac{\langle u_\mathcal{Q}^{m} | \Gamma^\nu | u_\mathcal{P}^{m+\lambda} \rangle \, f_{\nu}(\omega, \mu_\nu, \beta_\nu) \, d\omega}{(\omega - E_\mathcal{P}^{m+\lambda} - i\gamma_\mathcal{P})(\omega - E_\mathcal{Q}^{m} + i\gamma_\mathcal{Q})
}.
\label{eq:n_alpha_beta_q}
\end{equation}
Here, $E_{\mathcal{P}}^{m} = E_{\mathcal{P}} + m\omega$, and the Fermi-Dirac distribution is given by $f_{\nu}(x) = \left(1 + e^{x/\beta_\nu} \right)^{-1},$ where $\beta_\nu = 1/T$, $T$ denoting the temperature of the reservoir and $x$ denotes the energy relative to the chemical potential. After some algebraic calculations, it can be shown that for the identical reservoirs with chemical potential $\mu_R$ and in the weak coupling limit $\Gamma^\nu \to 0 $ (thus avoiding thermalization that may cause the loss of Floquet signatures and also ensuring mathematical simplicity via removing off-diagonal or time-dependent contributions), the Fourier components reduce to $n_{\mathcal{P}\mathcal{Q}}^{\lambda} \approx n_{\mathcal{P}} \, \delta_{\mathcal{P}\mathcal{Q}} \delta_{\lambda0},$
with the occupation probability given by
\begin{equation}
n_{\mathcal{P}}(\mu_R) = \sum_m \langle u_{\mathcal{P}}^{m} | u_{\mathcal{P}}^{m} \rangle \, f_R(E_{\mathcal{P}} + m\omega - \mu_R),
\label{eq:n_p_weak}
\end{equation}
where $f_R(E_{\mathcal{P}} + m \omega - \mu_R) $ is expressed as
\begin{eqnarray}
f_R(E_{\mathcal{P}}\!+\!m\omega\!-\!\mu_R)\!&=&\!\!\frac{i}{2\pi}\biggl[\Gamma_D\!\left(\!\frac{1}{2}\! + \!\frac{\beta_R\xi_{\mathcal{P}}^{\nu,m}}{2\pi} \!\right)\nonumber\\
&&-\Gamma_D\!\left(\!\frac{1}{2}\!-\!\frac{\beta_R\xi_{\mathcal{P}}^{\nu,m}}{2\pi} \!\right)\!-\!i\pi\biggr], 
\label{Fermi_function}
\end{eqnarray}
where $\xi_{\mathcal{P}/\mathcal{Q}}^{\nu,m} = i E_{\mathcal{P}/\mathcal{Q}}^{m} - i\mu_\nu,
$
and $\Gamma_D(\cdots)$ denotes the digamma function. We are now taking the limit $T \rightarrow 0$, or, $\beta_R \rightarrow \infty$. In this asymptotic limit, the digamma function can be expanded as
\begin{eqnarray}
\Gamma_D(z) \sim \ln z - \frac{1}{2z} - \sum_{n=1}^{\infty} \frac{B_{2n}}{2n z^{2n}}, \label{Gamma_Function}
\end{eqnarray}
Now substituting Eq.~(\ref{Gamma_Function}) in Eq.~(\ref{Fermi_function}) we get
\begin{eqnarray}
f_R = \frac{i}{2\pi} \left[ \text{ln} \left(\frac{\beta_R \, \xi_{\mathcal{P}}^{\nu,m}}{\pi} \right) - \text{ln}  \left( -\frac{ \beta_R \, \xi_{\mathcal{P}}^{\nu,m}}{\pi} \right) -i\pi  \right],
\end{eqnarray}
which reduces to $f_R = 1$ as $\text{ln} \left( -\frac{ \beta_R \, \xi_{\mathcal{P}}^{\nu,m}}{\pi} \right) = \left[\text{ln} \left( \frac{ \beta_R \,\xi_{\mathcal{P}}^{\nu, m}}{\pi} \right) + i\pi\right]$. Thus, Eq.~(\ref{eq:n_p_weak}) becomes $n_{\mathcal{P}} = 1$ for the normalized Floquet states. Hence, Eq.~\eqref{current} gives the simplified form of the current as
\begin{equation}
    \hat{I}(t)
    = \sum_{\mathcal{P}}  \langle u_{\mathcal{P}}(t)| \partial_{\phi}\hat{\mathcal{H}}(t) |u_{\mathcal{P}}(t)\rangle 
    \label{appenD2}.
\end{equation}
Using the relation $\mathcal{H}(t)|u_{\mathcal{P}}\rangle = (E_{\mathcal{P} }+i\partial_t) |u_{\mathcal{P}}\rangle$, the time-averaging of Eq.~\eqref{appenD2} yields
\begin{eqnarray}
I&=& \frac{1}{\mathcal{T}} \int_0^\mathcal{T} \sum_{\mathcal{P}}[\langle u_{\mathcal{P}}(t)| \partial_{\phi} (\hat{\mathcal{H}}(t) |u_{\mathcal{P}}(t)) \nonumber\\
&&-\langle u_{\mathcal{P}}(t)| \hat{\mathcal{H}}(t) | \partial_{\phi} u_{\mathcal{P}}(t) \rangle]=\sum_{\mathcal{P}}\partial_{\phi}E_{\mathcal{P}},    
\end{eqnarray}
which is the relation for the Josephson current in the driven scenario (Eq.~\eqref{floquet_current}).
\begin{figure*}
\centering
\includegraphics[width=0.75\linewidth]{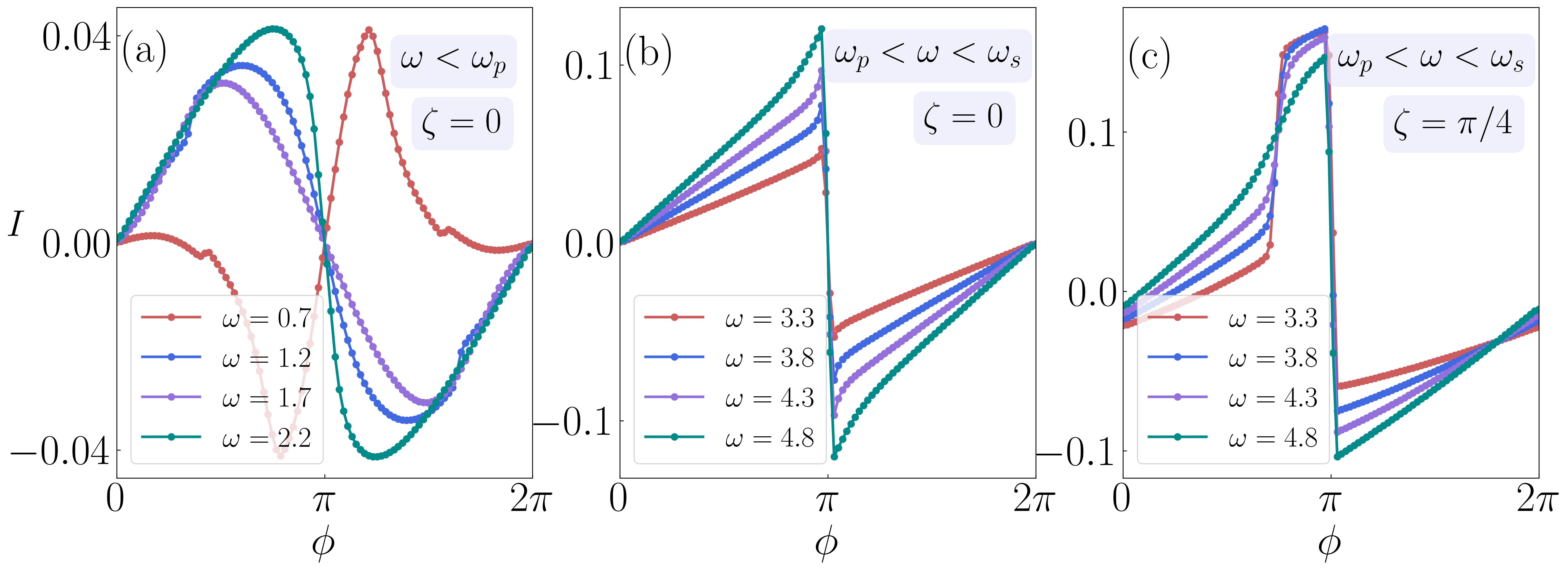}
\caption{CPR for different driving frequencies are presented for a fixed chemical potential \(\mu = 1\). (a) corresponds to the low-frequency regime (\(\omega < \omega_p\)), while (b) represents the intermediate-frequency regime (\(\omega_p < \omega <\omega_s\)), both in the absence of $\zeta$. (c) illustrates the CPR in presence of finite $\zeta$, namely, $\zeta=\pi/4$.
The rest of the parameters are fixed as those in Fig.~\ref{fig:static_E_vs_mu}.}
\label{fig:current_at_different_frequencies}
\end{figure*}
\begin{figure}
\includegraphics[width=\linewidth]{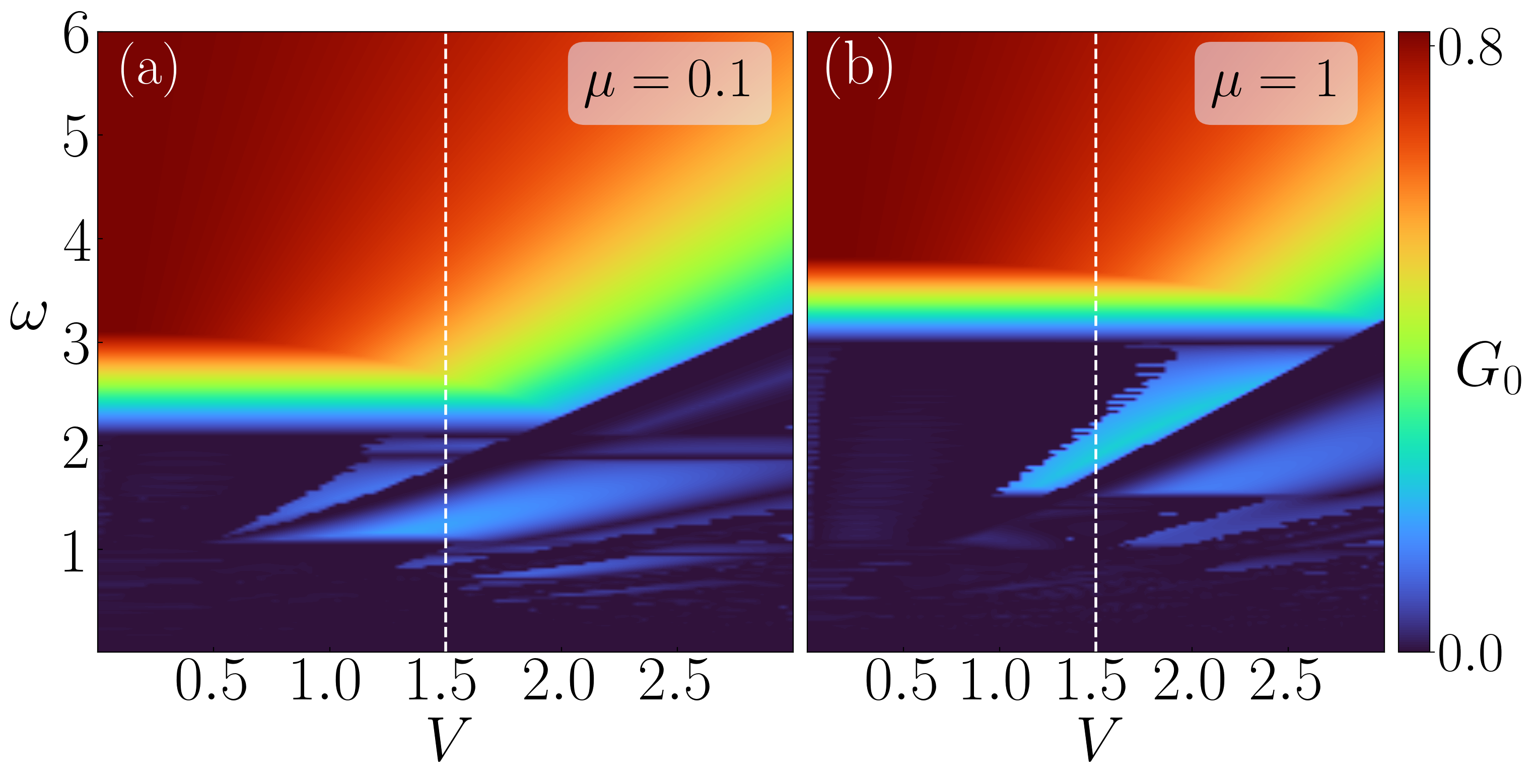}
\caption{The bulk gap invariant \(G_0\) is plotted in the \(V\)-\(\omega\) plane corresponding to two different values of chemical potential, namely, \(\mu = 0.1\) (panel (a)) and \(\mu = 1\) (panel (b)). The rest of the parameters are fixed as those in Fig.~\ref{fig:static_E_vs_mu}.}
\label{fig:gap_invariant}
\end{figure}
\section{Current and bulk gap at different driving frequencies}\label{appendixE}
Here, we would like to clarify the origin of the irregularities observed in the RF ($\mathcal{R}$) in the low-frequency regime, as shown in Fig. \ref{fig:RC_vs_omega}. As a dividend, this discussion will help in elucidates the linear dependence of the rectification factor in the intermediate-frequency regime as well. To begin with, we emphasize that the irregular behavior at low frequencies is not due to any numerical artefacts, but appears to be an intrinsic feature of the system. To support this claim, we have increased the number of time steps and observed that the irregular pattern persists, indicating that these features are physically grounded rather than numerical in origin.
\par We attribute this behavior to two key factors: (a) deviation from simple sinusoidal current-phase relations at low frequencies, and (b) suppression and fluctuation of the bulk gap at low frequencies. To gain deeper insight, we have plotted the CPR for $\mu = 1$, comparing both low-frequency ($\omega < \omega_p$) and intermediate-frequency (\(\omega_p<\omega< \omega_s\)) regimes, where the crossover frequency $\omega_p$ is identified from Fig. \ref{fig:RC_vs_omega}. Here, as shown in Fig. \ref{fig:current_at_different_frequencies}(a), the CPR in the low-frequency regime lacks a clear periodic structure, which contributes to the non-uniform behavior of $\mathcal{R}$. In contrast, above the crossover frequency $\omega_p$ (\ref{fig:current_at_different_frequencies}(b)), the system exhibits a well-defined (Majorana-mediated) $4\pi$-periodic CPR, with the amplitude increasing with frequency. This observation underpins the linear growth of $\mathcal{R}$ in the intermediate-frequency regime. On the other hand, \ref{fig:current_at_different_frequencies}(c) illustrates the CPR in presence of finite $\zeta$, namely, $\zeta=\pi/4$. Clearly, $I_-$ grows more rapidly in contrast to $I_+$, resulting in a net linear decline of $\mathcal{R}$, as observed in Fig. \ref{fig:RC_vs_omega}.

Additionally, the behavior of the rectification in different frequency regimes can also be understood by examining the bulk gap invariant ($G_0$) as a function of driving frequency, ($\omega$). To begin with, let us reconcile the fact that at very low frequencies, there occurs a squeezing of the Floquet Brillouin zone [$-\omega/2:\omega/2$], resulting in a suppressed
and a non-uniform bulk gap. Fig.~\ref{fig:gap_invariant} demonstrates this, showing the bulk gap in the $\omega$–$V$ plane. The vertical line at $V_1 = V_2 = V = 1.5$ indicates the specific regime where we ascertain the efficiency of our method and compare it with
the results of Fig.~\ref{fig:RC_vs_omega}. From this figure, it is evident that in the low-frequency regime, the system lacks a well-defined bulk gap. This not only undermines the protection of MZMs, but also reduces the accuracy of the energy derivative (used to
compute Josephson current) obtained numerically. Furthermore, we find that in the low-frequency
regime, the bulk gap exhibits significant non-uniformity. This irregular behavior  correlates well with the non-uniformity observed in the Josephson current (at low frequencies),
as discussed earlier (see Fig.~\ref{fig:current_at_different_frequencies}). These findings reinforce the conclusion that the non-uniform features have a genuine physical origin at low driving frequencies. Also, one can perform similar analyses for other values of the chemical potential to confirm this understanding.

\bibliography{bibfile}{}
\end{document}